\documentclass{aa}
\usepackage{graphicx}
\usepackage{supertab}
\usepackage{amssymb}
\begin{document}
%  \thesaurus{11(02.01.2; 11.01.2; 11.14.1; 11.19.1; 11.17.2)}
\title{Profile variability of the H$\alpha$ and H$\beta$ broad
emission lines in NGC~5548}
%\thanks{tables~\ref{tab2},\ref{tab5} are only available in electronic
%form at CDS via http://www.edsciences.org}

\author{A.I.\,Shapovalova\inst{1,5}
       \and V.T.\,Doroshenko\inst{2,7}
       \and N.G.\,Bochkarev\inst{2}
       \and A.N.\,Burenkov\inst{1,5}
       \and L.\,Carrasco\inst{3}
       \and V.H.\,Chavushyan\inst{3}
       \and S.\,Collin\inst{4}
       \and J.R.\,Vald{\'e}s\inst{3}
       \and N.\,Borisov\inst{1}
       \and A.-M.\,Dumont\inst{4}
       \and V.V.\,Vlasuyk\inst{1}
       \and I.\,Chillingarian\inst{2}
       \and I.S.\,Fioktistova\inst{1}
       \and O.M. \,Martinez\inst{6}
}
 \offprints{\\ A.I.~Shapovalova \email {ashap@sao.ru}}
 \institute{Special Astrophysical Observatory of the Russian AS,
Nizhnij Arkhyz, Karachaevo-Cherkesia, 369167, Russia
    \and Sternberg Astronomical Institute, University of Moscow,
Universitetskij Prospect 13, Moscow 119899, Russia
 \and
 Instituto Nacional de Astrof\'isica, Optica y Electr\'onica, INAOE,
Apartado Postal 51 y 216, 7200, Puebla, Pue., M\'exico
\and
 LUTH, Observatoire de Paris, Section de Meudon, Place Janssen,
92195, Meudon
France
\and
 Isaac Newton Institute of Chile, SAO Branch, Russia
\and Benem\'erita Universidad Aut\'onoma de Puebla, Facultad de
Ciencias F\'{\i}sico-Matem\'aticas, Apdo. Postal 1152, C.P.
72000, Puebla, Pue. M\'exico
\and
 Isaac Newton Institute of Chile, Crimean Branch, Ukraine
}

\date  {Received: 10 November 2003 / Accepted: 26 April 2004}
\titlerunning {H$\beta$ H$\alpha$ monitoring of NGC 5548}
\authorrunning{Shapovalova, A.I., et al.}

\abstract{ Between  1996 and 2002, we have carried out a spectral
monitoring program for the Seyfert galaxy \object{NGC 5548} with
the 6 m and 1 m telescopes of SAO (Russia) and with the 2.1 m
telescope of Guillermo Haro Observatory (GHO) at Cananea,
M\'exico. High quality spectra with S/N$>50$ in the continuum
near H$\alpha$ and H$\beta$ were obtained, covering the spectral
range $\sim$(4000 -- 7500)\,\AA \/ with a (4.5 to 15)\,
\AA-resolution. We found that both the flux in the lines and the
continuum gradually decreased, reaching minimum values during
May-June 2002. In the minimum state, the wings of H$\beta$ and
H$\alpha$ became extremely weak, corresponding to a Sy1.8 type,
not to a Sy1, as observed previously when the nucleus was
brighter. The line profiles were decomposed into variable and
constant components. The variable broad  component is well
correlated with the continuum variation. It consists of a double
peaked structure with radial velocities $\sim  \pm$1000~km/s
relative to the narrow component. A constant component, whose
presence is independent of the continuum flux variations, shows
only narrow emission lines.The mean, rms, and the averaged over
years, observed and difference line profiles of H$\beta$ and
H$\alpha$ reveal the same double peaked structure. The relative
intensity of these peaks changes with time. During 1996, the red
peak was the brightest, while in 1998 -- 2002, the blue peak
became the brighter one. Their radial velocities vary in the
$\sim$ (500 -- 1200)~km/s range. In 2000 -- 2002 a distinct third
peak appeared in the red wing of H$\alpha$ and H$\beta$ line
profiles. The radial velocity of this feature decreased between
2000 and 2002: from the observed profiles, from $\sim$ +(2500 --
2600)\, km/s to $\sim +2000$\, km/s and is clearly seen on the
difference profiles. The fluxes of the various parts of the line
profiles are well correlated with each other and also with the
continuum flux. The blue and red parts of the line profiles at
the same radial velocities vary in an almost identical manner.
Shape changes of the different parts of the broad line are not
correlated with continuum variations and, apparently, are not
related to reverberation effects. Changes of the integral Balmer
decrement are, on average, anticorrelated with the continuum flux
variations. This is probably due to an increasing role of
collisional excitation as the ionizing flux decreases.
 The behavior of the
Balmer decrement of the various parts of the line profiles was
different in 1996 -- 2000 as compared with the 2001 behavior. Our
results favor the formation of the broad Balmer lines
  in a turbulent
 accretion disc with large and moving
"optically thick" inhomogeneities, capable of reprocessing the central source continuum.
 }

\maketitle

\section{ Introduction}
An important question in the study of active galactic nuclei
(AGN) is the nature of the ``central engine''. A popular
assumption is that the nuclear  activity is caused by accretion
of gas onto a supermassive black hole (Rees~\cite{ree},
Begelman~\cite{beg}). The basic energy release of an AGN occurs
very close to the nucleus ($r<0.001$ pc), as an UV and X-ray
continuum most probably produced by a geometrically thin accretion
 disc. Then,
broad emission lines are produced in a zone (the BLR) that reprocesses a
fraction of the central UV-X continuum. A zone located
further out ($r>0.001$ pc). The BLR is filled with gas
obviously linked with the accretion process. It is therefore
important to know its structure and kinematics, in order to gain
insight into the central engine. However, even for nearby
objects, the typical angular size of the BLR corresponds to  $<
0.001$ arcsec, hence we will have to wait for the availability of
more sensitive optical interferometers to resolve it.
Fortunately, another methods exist to study the BLR structure.

It is well known that AGNs vary in luminosity on
time scales from years to hours, over the entire wavelength range
 from
radio to X-ray or $\gamma$-ray. In particular, the flux in the broad emission
lines varies in response to changes in the ionizing continuum
with short time delays (days to weeks for Seyfert galaxies), due
to light-travel time effects within the BLR. If the BLR gas has
systematic motions such as infalling, outflowing, circular
motions, etc., then the profiles of the broad emission lines must
vary in a way related with the geometry and the kinematics of the gas
in
this region, and with the processes of gas relaxation that follow
the changes in the ionizing flux (Bahcall, Kozlovsky \& Salpeter 1972; Bochkarev \&
Antokhin~\cite{boc}; Blandford \& McKee~\cite{bla}; Antokhin \&
Bochkarev~\cite{ant}).

Studying the correlations between the flux changes in the
continuum and in the broad emission line profiles,  one can
obtain a ``map'' of the geometrical and dynamical structure of
the BLR. This method is known as ``Reverberation mapping''(see
Peterson~\cite{pet1993} and references therein). Important
progress in the understanding of the BLRs was achieved as a
result of multiwavelength monitoring campaigns within the
framework of the  ``International AGN Watch'', a consortium
organized to study several Seyfert galaxies (Peterson et al.
~\cite{pet1999}). A large amount of data has been obtained in the
multiwavelength monitoring of the Seyfert 1
 galaxy NGC 5548, including its continuous monitoring in the
optical range for 13 years (1988 -- 2001) (Clavel et
al.~\cite{cla}; Peterson et al.
~\cite{pet1991,pet1994,pet1999,pet2002}; Dietrich et
al.~\cite{die1993}; Korista et al.~\cite{kor1995}). These
investigations have given the following results:
\begin{enumerate}
\item
The response time of the H$\beta$ line  to
continuum
variations varies from year to year (from $\sim$ 8 to 26 days) and
it is
correlated with the average continuum flux.
\item
The high ionization lines have shorter response times to continuum
variations
than the low ionization ones, indicating the presence of ionization
stratification
along the radius of the broad line region.
\item
The optical and ultraviolet continua vary almost simultaneously (i.e.
without showing a significant delay).
\item
The UV/optical continuum becomes ``harder'' as it gets brighter.
\end{enumerate}
The results of the study of the radial velocity field in
the BLR of NGC 5548 are still ambiguous. The analysis  of the
``AGN Watch'' spectra reveals a complex behaviour. Crenshaw and
Blackwell (\cite{cre}) found that the red wing of
\ion{C}{iv}$\lambda$\,1550 responds faster to changes of
the continuum than the blue wing does, implying important radial
motions (infalling). Later on, Korista (\cite{kor1994}) observed
that both wings presented the same delay with
respect to continuum light variations, implying Keplerian
rotation or turbulent symmetrical motions. The absence of
important radial motions in the BLR was further confirmed by the
analysis of the transfer function (TF)  of the H$\beta$ (Wanders
\& Peterson ~\cite{wan1996}) and \ion{C}{iv}$\lambda$\,1550
(Wanders et al.~\cite{wan1995}) emission lines. It was also found
that the broad wings of the emission lines respond faster to
continuum variations than the line cores  (Clavel et
al.~\cite{cla}; Korista~\cite{kor1994}; Korista et
al.~\cite{kor1995}; Kollatschny \& Dietrich~\cite{kol},  Wanders
\& Peterson~\cite{wan1996}). In several studies, the presence
of a multi-component structure of the BLR with distinct physical
characteristics is required to account for the  observed broad
line profile variability in this object
(Peterson~\cite{pet1987a}, Stirpe et al.~\cite{sti1988}; Stirpe \&
de Bruyn~\cite{sti1991}; Sergeev et al.~\cite{ser}; Wanders and
Peterson~\cite{wan1996}). Consequentely, the observed profiles in NGC 5548 have
been interpreted in the framework of different models, such as: binary
black holes (Peterson et al.~\cite{pet1987b}), accretion disc
(Stirpe et al.~\cite{sti1988}, Rokaki \& Boisson~\cite{rok}), BLR
from the clouds rotating about the massive central gravitational
source in occasionally inclined Keplerian orbits and illuminated
by an anisotropic continuum source (Wanders et al.~\cite{wan1995},
Wanders \& Peterson~\cite{wan1996}).

Further study of the broad emission line profile changes on longer
time scales, may allow us to prove or disprove some of the
previously advanced hypothetical scenarios. This is the purpose of
the present paper.

In this paper we present the results of an optical spectral study
of NGC~5548 for the 1996 -- 2001 period, including
part of our 2002 data (see section~\ref{sec2.1}). Some partial
results of our monitoring campaign were
reported earlier (Shapovalova et
al.~\cite{sha2001a},~\cite{sha2001b},~\cite{sha2002}). In
section~\ref{sec2}, we discuss the observations and data
processing. In section~\ref{sec3} we present the analysis of the
H$\beta$ and H$\alpha$ line profile variability. The
decomposition of the profiles into constant and variable
components, the mean and rms spectra, are discussed. The
behaviour of the radial velocities of spectral features in the broad
line profiles, is studied both from the observed profiles and from the
difference profiles. We use words "bump" for wide features
and "peak" for their tops (maxima).
 The correlation analysis between fluxes
 and shapes
of the different parts of the line profiles is presented.  The
behaviour of the
Balmer decrement is studied.
In Section~\ref{sec4}, we summarize our results and compare them with
those of
other researchers. Possible interpretations are discussed
 in Section~\ref{sec5}, and conclusions are listed in
 Section~\ref{sec6}. In Appendix~\ref{ap}, a modified spectrum scaling
  method adopted in this paper is presented.

    Note that all these data are also included in the AGN Watch database
and are publicly available.

\section{ Observations and data reduction}
\label{sec2}
\subsection{ Optical Spectroscopy}\label{sec2.1}

We report the spectral observations of NGC 5548 carried out
between 1996 Jan 14 (Julian date= JD 2\,450\,097) and 2001 Aug 9
(JD 2\,452\,131) during 113 nights. Our analysis
is based on those spectra. However for the studies described in
Sections~\ref{sec3.1}, \ref{sec3.2.3} and \ref{sec3.2.4}
(for studing the behavior of the peaks and bumps) we have also used spectra
taken on June 4, 2002, and May 15, 17, 2002, when NGC~5548 was in
a minimum activity state and the 2002 annual averages of both the
observed and difference profiles for H$\alpha$ and
H$\beta$.

Optical spectra of NGC 5548 were obtained with the 6 m and 1 m
telescopes of  SAO  (Russia, 1996 -- 2002) and at INAOE's 2.1 m
telescope at the Guillermo Haro Observatory (GHO) at Cananea,
Sonora, M\'exico (1998 -- 2002). These were obtained with
long slit spectrographs equipped with CCDs. The typical wavelength
range covered was from 4000\,\AA\, to 7500\,\AA, the spectral
resolution was 4.5 -- 15\,\AA, and the S/N ratio was $>50$ in the
continuum near H$\alpha$ and H$\beta$. Spectrophotometric
standard stars were observed every night. The informations on the
source of spectroscopic observations are given in
Table~\ref{tab1}: 1 - the source (Observatory); 2 - a code assigned to each telescope+equipment,
 used throughout this
paper (the code was chosen in accordance with the monitoring
campaigns of NGC 5548 (Dietrich et al.~\cite{die2001}; Peterson
et al.~\cite{pet2002}); 3 - the telescope aperture  and the
spectrograph; 4 - the projected spectrograph entrance apertures
(the first dimension is the slit-width, and the second one is the
slit-length).

\begin{table}
\begin{center}
\caption{Sources of Spectroscopic Observations}
\label{tab1}
\begin{tabular}{llll}
\hline
\hline
Source& Code& Tel.and Equip.& Aperture\\
\hline
1&2& 3 &4\\
\hline
SAO (Russia)&    L1&  1m+UAGS &   $4.2''\times 19.8''$\\
SAO (Russia)&    L1&  1m+UAGS &   $8.0''\times 19.8''$\\
SAO (Russia)&    L &  6m+UAGS &   $2.0''\times 6.0''$\\
Guillermo Haro &  GH&  2.1m+B\&C&   $2.5''\times 6.0''$\\
Obs. (Mexico)    &       &         &             \\
\hline
\end{tabular}
\end{center}
\end{table}
%%%%%%%%%%%%%%%%%%%%%%%%%%%%%%%%%%%%%%%%%%%%%%%%%%%%%%%
\begin{table}[h ]
\caption{Log of the spectroscopic observations} \label{tab2}

This table is only available in electronic form at the CDS. It
contains the following information for 116 dates: Columns: 1 -
number; 2 - UT date; 3 - Julian date; 4- code according to Table
1; 5 - projected spectrograph entrance apertures; 6 - wavelength
range covered; 7 - spectral resolution; 8 - mean seeing; 9 -
position angle (PA) in degrees; 10 - signal to noise ratio in the
continuum (5160 -- 5220)\,\AA\AA\, near H$\beta$ and (6940 --
7040)\,\AA\AA\, near H$\alpha$.

\end{table}
%%%%%%%%%%%%%%%%%%%%%%%%%%%%%%%%%%%%%%%%%%%%%%%%%%%%%%%%
The spectrophotometric data reduction was carried out either
with the
software developed at the SAO RAS by Vlasyuk (\cite{vla}), or with
IRAF for
the spectra obtained in M\'exico. The image  reduction process
included bias
subtraction, flat-field corrections, cosmic ray removal, 2D wavelength
linearization, sky spectrum subtraction, stacking of the spectra for
every
set-up, and  flux calibration based on standard star observations.

\subsection{Absolute calibration of the spectra}
\label{sec2.2}
 Even under good photometric conditions, the
accuracy of spectrophotometric measurements is  rarely better
than 10\%. Thus the standard
technique of flux calibration, by means of comparison with stars
of known spectral energy distribution, is not good enough
 for the study of AGN variability.
Instead, we use the fluxes of the narrow emission lines which are
known to be non-variable on time scales of tens of years in most
AGN. Consequently, the bright narrow emission lines can be
adopted as internal calibrators for scaling AGN spectra
(Peterson~\cite{pet1993}). So, we assume that the flux of the
[\ion{O}{iii}]$\lambda$\,5007 line remains constant during the
interval covered by our observations. All blue spectra of NGC 5548
are scaled to a constant flux  value of $
F$([\ion{O}{iii}]$\lambda\,5007)= 5.58\times
10^{-13}$\,erg\,s$^{-1}$\,cm$^{-2}$  determined by Peterson et
al. (\cite{pet1991}) and corrected for aperture effects as
described below. The scaling of the blue spectra was carried out
using a variation on the method of Van Groningen \& Wanders
(\cite{van}), described in Appendix\,\ref{ap}. This method allows
to obtain a homogeneous set of spectra with the same wavelength
calibration and the same [OIII]~$\lambda$5007 flux value.

The spectra obtained with the GHO 2.1~m telescope (M\'exico) with
a resolution of 15\AA. They contain  both the H$\alpha$ and
H$\beta$ regions, and were scaled using the
[\ion{O}{iii}]$\lambda$\,5007 line.

Most of spectra from the 1~m and 6~m SAO telescopes were obtained
separately in
the blue (H$\beta$) and red (H$\alpha$) wavelength intervals,
 with a resolution of 8 -- 9\,\AA. Usually, the red edge
of the blue spectra and the blue edge of the red spectra
overlap in an interval of $\sim 300$\,\AA. Therefore, the
most of the red spectra were scaled using the overlapping
continuum region with the blue ones, which were scaled with the
[\ion{O}{iii}] line. In these cases the scaling
uncertainty is about 5\%.

However, for 10 red spectra the scaling of the continuum by this
method was not possible, this due to several reasons: ie. some spectra
were obtained with a
higher resolution ($\sim5$\AA) and did not overlap with the blue
spectra; or the some spectrum ends were distorted by the
reduction procedure; or blue
spectra were not taken on that night. These 10 spectra were
scaled using the integral flux in the narrow emission lines in
the H$\alpha$, region: [\ion{N}{ii}]$\lambda\lambda$\, 6548,\,
6584 and [\ion{S}{ii}]$\lambda\lambda$\, 6717,\, 6731. To this
purpose, on the red spectra we located a linear continuum through
points that are free from the absorption lines (6120\,\AA\, and 7020\,\AA)
in 20\,\AA windows.

\begin{figure}
\centering
\includegraphics[width=9.5cm,angle=-90,bb=52 106 540 521,clip=]{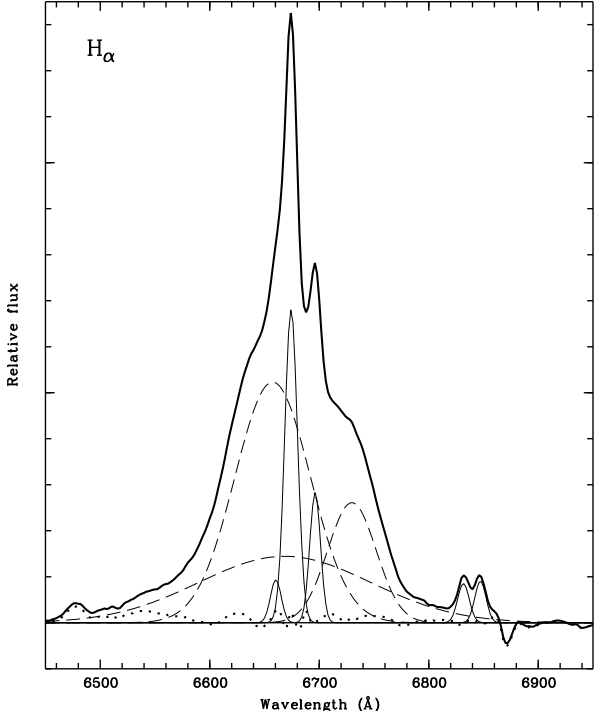} \caption{
  Gaussian deconvolution of the H$\alpha$ blend in several
  components;
  thick solid line: the observed profile;
thin solid line: the narrow component of  H$\alpha$,
[\ion{N}{ii}]$ \lambda\lambda$\,6548, 6584,
[\ion{S}{ii}]$\lambda\lambda$\,6717, 6731; dashed line: very
broad, blue broad, and red broad components of H$\alpha$; points:
the residual of the observed profile and the sum of all
gaussian components. Our analysis do not include a small hump in
the left side ([\ion{O}{i}]$ \lambda$\,6364).}
\label{fig1}
\end{figure}

After continuum subtraction, we obtained the best gaussian fit to
the H$\alpha$ profile through a blend of 7 emission components:
very broad H$\alpha$ (FWHM$\sim 10000$ km/s), blue broad
H$\alpha$, red broad  H$\alpha$, narrow H$\alpha$,
[\ion{N}{ii}]$\lambda\lambda$\,6548+6583 (double gaussian
function with I(6584)/I(6548)=3), [\ion{S}{ii}]$\lambda$\,6717 and
[\ion{S}{ii}]$\lambda$\,6731. Fig.~\ref{fig1} shows an example of
a Gaussian fit to the H$\alpha$ blend.

 Table~\ref{tab3} lists our mean values of the flux in the
narrow components derived from the scaled red spectra, jointly
with the result obtained by Dietrich et al. (\cite{die1993}) from
the spectra taken in 1988 -- 1989. These authors obtained a good
Gaussian fit to H$\alpha$ with only 5 components: namely, very
broad H$\alpha$, broad H$\alpha$, narrow H$\alpha$,
[\ion{N}{ii}]$\lambda$$\lambda$\,6548,\, 6584. With our spectra
we tried to carry out a similar fitting to the same 5 components,
but we ended up with large residuals ($\sim$ 10 -- 20\%) in the
region  where the blue and red bumps appear. Apparently, on the
spectra in Dietrich et al. (\cite{die1993}), such features were
not as bright as in our spectra. However, in any profile fit, the
narrow components should remain constant.

\begin{table}
\small
\caption{Fluxes of H$\alpha$  Gaussian components}
\label{tab3}
\begin{tabular}{l|ll}
\hline
\hline
  Component&\multicolumn{2}{c}{  Flux*}\\
&                     mean (GHO+SAO)& Dietrich et al. \\
&&(\cite{die1993})\\
\hline

 H$\alpha_{\rm narrow}$  &      $0.350\pm 0.040$&   $0.344\pm 0.045$\\

[\ion{N}{ii}]$\lambda\lambda$\,6548, 6584&  $0.160\pm 0.036$&
$0.182\pm 0.022$\\

[\ion{S}{ii}]$\lambda\lambda$\,6717, 6731&        $0.097\pm 0.021$&\\
Sum (H$_{\alpha {\rm n}}$, [\ion{N}{ii}])&
$0.510\pm0.054$&$0.526\pm0.050$\\

Sum narr. all&     $0.607\pm 0.064$&\\

\hline
\multicolumn{3}{l}{*\rule{0pt}{11pt}$F$([\ion{O}{iii}]$\lambda$\,5007)=1.0}\\
\end{tabular}
\end{table}

We can see on Table~\ref{tab3} that the fluxes obtained for the
narrow lines are consistent within the errors with the values
obtained by
Dietrich et al. (\cite{die1993}). Using the integral fluxes for
the narrow lines from Table~\ref{tab3} we scaled the 10 red
spectra.

\subsection{Intercalibration of the spectral data}
\label{sec2.3}

>From the scaled spectra we determined the average flux in the
continuum at the observed wavelength $\sim 5190$\AA\, (or at $\sim
5100$\AA\, in the rest frame of NGC 5548, z=0.0167), by means of
flux averages in the bandpass (5180 -- 5200)\,\AA. For the
determination of the observed fluxes of H$\beta$ and H$\alpha$,
it is necessary to subtract the continuum. To this goal, a linear
continuum was located through windows of 20\,\AA\,   located at
4790\,\AA\, and 5170\,\AA\, for the H$\beta$ region, and at
6120\,\AA\, and 7020\,\AA\, for the H$\alpha$ region. After the
continuum subtraction, we defined the observed fluxes in  the
lines in the following wavelength intervals: (4795 -- 5018)\,\AA\,
for H$\beta$ (the interval is similar to that in  Peterson et al.
\cite{pet2002}), and (6500 -- 6800)\,\AA\, for H$\alpha$ (the
interval is like that in Dietrich et al. \cite{die2001}).

All fluxes were corrected for aperture effects because, while the
BLR and non-stellar continuum are effectively point-like sources
($\ll 1''$), the NLR is an extended one ($>2''$). Consequently,
the measured NLR flux depends on the size of the spectrograph's
entrance aperture (see Peterson et al. \cite{pet1995}, for a
detailed discussion). In order to correct our fluxes for aperture
effects, we determined  a point-source correction factor
($\varphi$) given by:

\begin{equation}
{\rm F(H\beta)=\varphi\cdot F_{\lambda5007}\left(\frac{F(H\beta)}
{F([\ion{O}{iii}]\lambda5007)}\right)_{obs}} ,
\label{equ1}
\end{equation}
where $F_{\lambda5007}$ is the absolute flux in the
[\ion{O}{iii}]$\lambda$\,5007 line, and the value in brackets is the
H$\beta$
to [\ion{O}{iii}]$\lambda$\,5007 observed flux ratio.

The light contribution of the host galaxy to the continuum depends
also on
the aperture size. The continuum fluxes $ F_\lambda$(5190) were
corrected
for different amounts of host-galaxy contamination, according to
the following expression (see Peterson et al. \cite{pet1995}):

\begin{equation}
{\rm F_{\lambda}(5190\AA) = \varphi\cdot F_{\lambda
5007}\left(\frac{F_{\lambda}(5190\AA)}
{F([\ion{O}{iii}]\lambda 5007)}\right)_{obs}-G} ,
\end{equation}
where $F_{\lambda5007}$ is the absolute flux in the
[\ion{O}{iii}]$\lambda$\,5007 line and the value in brackets is
the continuum to [\ion{O}{iii}]$\lambda$\,5007 observed flux
ratio, $G$ being an aperture dependent correction factor to
account for the host galaxy light. The case L1
(Table~\ref{tab1}), which corresponds to a relatively large aperture
($4.2''\times 19.8''$), was adopted as a  standard (i.e.
$\varphi=1.0$, $G$=0 by definition). The corrections $\varphi$ and
$G$ were defined for every aperture via the comparison of a pair
of observations separated in time by 1-3 days. This  means that
the real variability on shorter times ($<3$ days) will be
suppressed by the procedure of data recalibration. Point-source
correction factors $\varphi$ and $G$ values for various samples
are given in Table~\ref{tab4}. Using these factors, we
recalibrate the observed fluxes in the lines and continuum to a
common scale corresponding to an aperture of $4.2''\times 19.8''$.

\begin{table}
\begin{center}
\caption{Flux Scale Factors for Optical Spectra}
\label{tab4}
\begin{tabular}{cccc}
\hline
\hline
  Sample  &   Aperture &    Point-Source & Extended Source\\
  &       & Scale factor & Correction \\
&&$\varphi$ &$G$*\\
\hline
    L1 &     $4.2''\times 19.8''$&   1.000 &          0.000\\
    L1 &     $8.0''\times 19.8''$&   $1.089\pm 0.031$&    $0.84\pm
0.45$\\
    GH &     $2.5''\times 6.0''$&    $1.009\pm 0.021$&   $-2.76\pm
0.11$\\
    L&       $2.0''\times 6.0''$&    $1.036\pm 0.048$&   $3.85\pm
0.86$\\
\hline \multicolumn{4}{l}{*\rule{0pt}{11pt}  in units of
$10^{-15}$\,erg\,s$^{-1}$\,cm$^{-2}$\,\AA$^{-1}$}\\
\end{tabular}
\end{center}
\end{table}

%%%%%%%%%%%%%%%%%%%%%%%%%%%%%%%%%%%%%%%%%%%%%%%%%%%%% TAB5
\begin{table}[h]
\caption{Observed continuum, H$\beta$ and H$\alpha$ fluxes.}
\label{tab5}

Table~\ref{tab5}, where we list our results of fluxes
measurement, is only available in electronic form at the CDS and
contains the following information. Columns: 1 - UT-Date; 2 -
Julian date; 3\,- a telescope code, according to
Table~\ref{tab1}; 4 - $F$(cont), the continuum flux at 5190\,\AA\,
(in units of $10^{-15}$\, erg\,s$^{-1}$\,cm$^{-2}$\,\AA$^{-1}$),
reduced to the 1 m telescope aperture $4.2''\times 19.8''$; 5 -
$\varepsilon_c$, the estimated  continuum flux error; 6 -
$F$(H$\beta$), the H$\beta$ total flux (in units of
10$^{-13}$\,erg\,s$^{-1}$\,cm$^{-2}$); 7 - $\varepsilon_{{\rm
H}\beta}$, the H$\beta$ flux error; 8\,- $F$(H$\alpha$), the
H$\alpha$ total flux (in units of
$10^{-13}$\,erg\,s$^{-1}$\,cm$^{-2}$); 9 -  $\varepsilon_{{\rm
H}\alpha}$, the H$\alpha$ flux error.
\end{table}
%%%%%%%%%%%%%%%%%%%%%%%%%%%%%%%%%%%%%%%%%%%%%%%%%%%%%%

The fluxes listed in Table~\ref{tab5} were not corrected for
the contributions of the narrow-line emission components of H$\beta$,
H$\alpha$, and [\ion{N}{ii}]$\lambda\lambda$\,6548, 6584. These
are constant and should not influence a broad line variability study. The mean error
(uncertainty) in our flux determination for both,
the H$\beta$ and the continuum, is $\sim 3$\%, while it is $\sim
5$\% for H$\alpha$. These quantities were estimated by comparing
our  results from the spectra obtained within time intervals
shorter than 3 days.

\subsection{Subtraction of the narrow emission line contribution}
\label{sec2.4}
In order to study the broad components of hydrogen
lines showing the main BLR characteristics, one must remove the
narrow component of these lines and the forbidden lines from the
spectra. To this purpose, we constructed spectral templates for
the H$\beta$ and the H$\alpha$ blends using a Gaussian fit to the
higher spectral resolution ($\sim 8$\,\AA) profiles observed near
the minimum light state (for details see section~\ref{sec2.2} and
Fig.~\ref{fig1}). The obtained template spectra are the sum of
the following gaussian components: for H$\beta$, the narrow
component of H$\beta$, [\ion{O}{iii}]$\lambda\lambda$\,4959, 5007;
for H$\alpha$, the narrow component of H$\alpha$,
[\ion{N}{ii}]$\lambda\lambda$\,6548, 6584 and
[\ion{S}{ii}]$\lambda\lambda$\,6717, 6731. The flux values obtained for
the narrow components of  H$\beta$ and H$\alpha$ in the template
spectra are:  $F({\rm H}\beta)_{n}=(0.131\pm
0.012)F(\lambda5007)$; $F({\rm H}\alpha)_{n}=(0.398\pm
0.02)F(\lambda5007)$, respectively. Then, we scaled the blue and
red spectra according to our scaling scheme (see
Appendix\,\ref{ap}), taking the template spectrum as a reference.
Our template spectrum and any individual observed
spectrum are thus matched in wavelength, reduced to the same
resolution, and then subtracted from one another. After subtraction
of the narrow components,
the spectra of the H$\alpha$ and  H$\beta$ broad lines is
reduced to the aperture $4.2''\times 19.8''$
(equation~\ref{equ1}), using  $\varphi$ values listed in
Table~\ref{tab4} (section~\ref{sec2.3}).

 Since the scaling is done from the line
[\ion{O}{iii}]$\lambda$\,5007, we had to use the FWHMs
 obtained from the corresponding blue spectrum to reduce
the  H$\alpha$ region. But the FWHM of
[\ion{N}{ii}]$\lambda\lambda$\,6548, 6584 and
[\ion{S}{ii}]$\lambda\lambda$\,6717, 6731 lines is usually
somewhat smaller (by $\sim$ 5 -- 10\%) than the FWHM of
[\ion{O}{iii}]$\lambda$\,5007. Therefore, the subtraction of
these components is rather poor when one has large spectral
resolution differences between the template and individual
spectra.

Another way to remove the narrow emission lines, consists in
subtracting the spectra from each other, i.e. obtaining
difference spectra. The H$\beta$ and H$\alpha$ difference
profiles are thus obtained by
subtracting the minimum activity state spectrum
from the individual spectra.
For a good subtraction, it is necessary for the spectra to have similar
spectral resolution.  These questions are
solved via the method mention in Appendix~\ref{ap}, where a spectrum
in minimum state is used as a reference spectrum. The difference
spectra will be discussed hereinafter.

\section{Data analysis}
\label{sec3}
\subsection{ Variability of H$\beta$ and H$\alpha$ emission lines
and of the optical continuum} \label{sec3.1}

\begin{figure}
\centering
\includegraphics[angle=-90,width=8.5cm,bb=50 60 570
770,clip=]{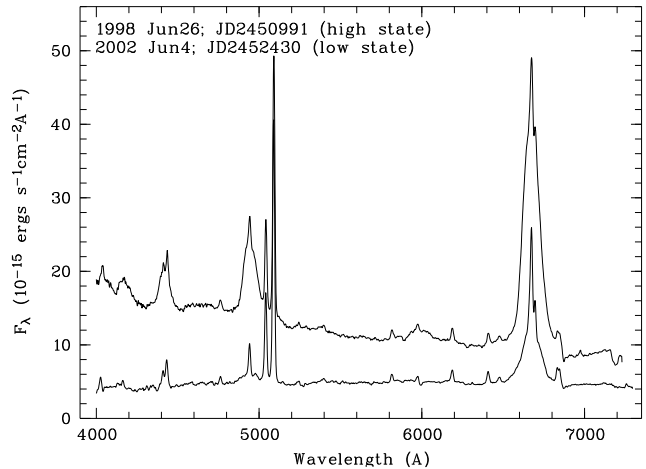} \caption{
  The spectra of NGC 5548 corresponding to the high activity state
  (top) and to the low activity state (bottom).
   The absissae give the
observed wavelengths
  (z=0.0167). The ordinates give the fluxes in units of
$10^{-15}$\,erg\,cm$^{-2}$\,s$^{-1}$\,\AA$^{-1}$.} \label{fig2}
\end{figure}

In Table~\ref{tab5} it is apparent that both, the continuum and
permitted line emission fluxes, decreased continuously from
maximum values in 1998 to minimum ones in 2002. The maximum
amplitude ratios of the flux variations during this period were:
for H$\beta$ line - $\sim 4.7$; for H$\alpha$ - $\sim3.4$; and
for the $\lambda \lambda 5190\,\AA $ continuum  - $\sim 2.5$.
Photometric data by Doroshenko et al.(~\cite{dor}) and Spiridonova
O.I.(~\cite{spi}) also yield a maximum amplitude of the flux
variations ratio $\sim 2.5$ in the V band in this period.
Thus, the variations inferred from the spectral and broad band
photometry of the continuum are in excellent agreement. This fact
is in turn indicative of a constant flux in the line
[\ion{O}{iii}]$\lambda$\, 5007, a key assumption in our spectral
scaling scheme. It is worth mentioning, that the galactic bulge
light contribution to our spectra was not subtracted. Therefore,
the derived maximum amplitude ratio for the continuum flux
changes is, consequently, smaller than the ones derived for
emission lines. In 2002, the flux in the lines and continuum
reached a minimum value by mid May or early June. In
fig.~\ref{fig2} we plot spectra of the high  and low activity
states, obtained in June 26 1998, and June 4 2002, respectively.
There one can see that in the low activity state, the flux in the
continuum decreased by a large factor ($\sim 2.5$ times), while
the slope of the continuum in the blue, became significantly
flatter, showing a spectral index $>2$ to be compared with the
$\sim1.0$ a value observed at the high activity state. Also, at
minimum activity state, the emission wings of H$\beta$ and
H$\alpha$  became extremely weak. These profiles correspond to a
Sy1.8 type and not to a Sy1, as observed in maximum light (i.e.
the spectral type of the object had suffered a dramatic change!).

\begin{figure}
%\resizebox{\hsize}{!}{\includegraphics[ bb=72 188 525 692]{0652f3.ps}} \vspace{0 mm}
\includegraphics[width=8.5cm]{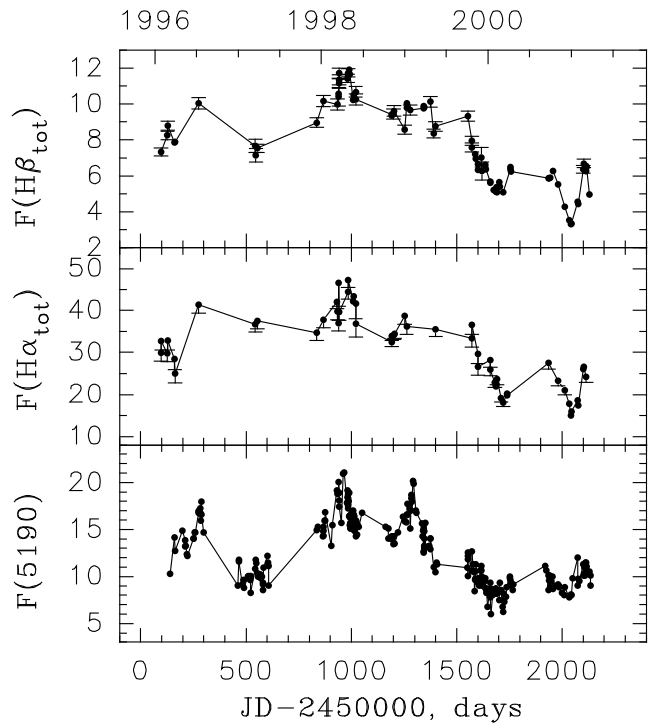}
\caption{The light curves of H$\beta$, H$\alpha$, and the
continuum at the observed wavelength $\lambda$5190\,\AA. The
narrow components were not subtracted. Fluxes in lines are in
units of $10^{-13}$\,erg\,cm$^{-2}$\,s$^{-1}$, and in continuum in
$10^{-15}$\,erg\,cm$^{-2}$\,s$^{-1}$\,\AA$^{-1}$. The continuum is
the ``combined continuum" obtained from the spectra and
photometric points in the V band in 1996 -- 2001} \label{lc}
\end{figure}

In Fig.~\ref{lc} we present the light curves of the H$\beta$ and
the H$\alpha$ emission lines and the combined continuum. The flux
in the lines and spectral continuum are those listed in
Table~\ref{tab5}. The combined continuum data includes both the
$\lambda$5190\, spectral data and some V band photometry from
Doroshenko et al. (\cite{dor}) and Spiridonova (\cite{spi}). The V
band data were converted into $\lambda$5190\ continuum flux
values through the expression given by Dietrich et al.
(\cite{die2001}):
\begin{equation}
{\rm log}F_{\lambda}(5500)=-0.4\,m_V-8.439.
\end{equation}
A comparison of the spectral continuum at $\lambda$5190\,\AA,
$F_c$($\lambda$5190) with some simultaneous observations made in
the V band leads to the following transformation equation:
\begin{eqnarray}
F_c (\lambda 5190)=(1.120\pm 0.064)\times F_{\lambda}(V) \nonumber \\
-(0.608\pm 0.776),
\end{eqnarray}
(correlation coefficient r$=0.95$). Here the fluxes are given in
units of 10$^{-15}$ erg~cm$^{-2}$\,s$^{-1}$\,\AA$^{-1}$.
One sees on Fig. 3  that the  H$\beta$ and H$\alpha$ fluxes
 vary approximately in the same way.
We have compared our light curves
of H-beta and of the continuum with those obtained by Peterson et al.
 (\cite{pet2002}) from the AGN-Watch program.
The comparison shows that our H-beta fluxes coincide within
the uncertainties with the AGNW data. The continuum fluxes differ because
of the different apertures, which
are $4.2''\times 19.8''$ for our data
and $5.0''\times 7.5''$ for the AGNW data.
The correction for the different apertures necessary to convert our
data to Peterson et al.'s data (\cite{pet2002}) is
F(cnt)pet.=F(cnt)our -2.5 (the fluxes are in units
10$^{-15}$
erg~cm$^{-2}$\,s$^{-1}$\,\AA$^{-1}$). With the introduction of such a correction,
our continuum fluxes
coincide with the AGNW data within the uncertainties.
%%%%%%%%%%%%%%%%%%%%%%%%%%%%%%%%    fig2

%%%%%%%%%%%%%%%%%%%%%%%%%%%%%%%%    fig3

\subsection{ H$\alpha$ and H$\beta$ profile variations}
\label{sec3.2}
\subsubsection{ H$\beta$ and H$\alpha$ profile decomposition}
\label{sec3.2.1}

\begin{figure}
\resizebox{\hsize}{!}{\includegraphics{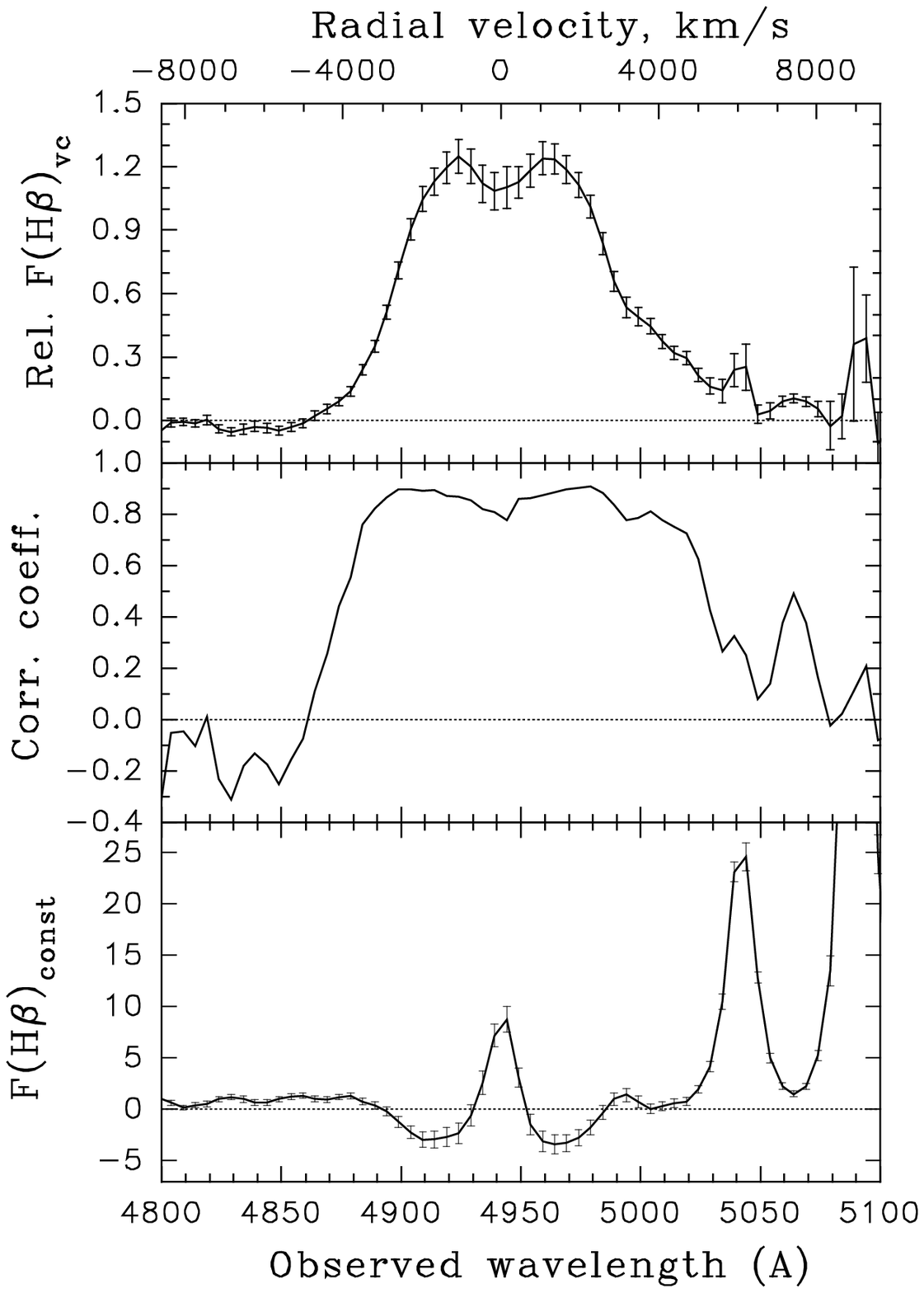}
\includegraphics{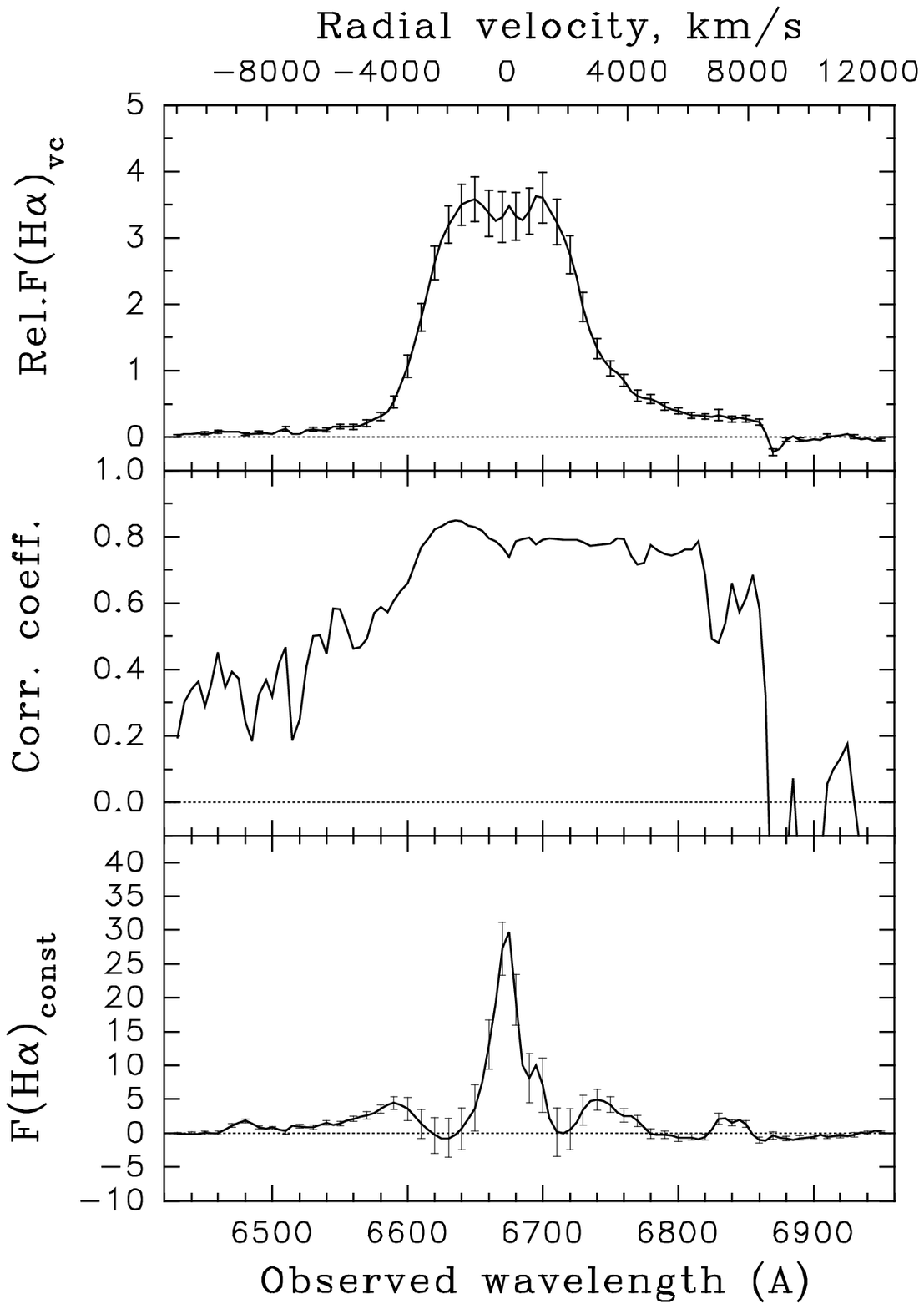}}
\caption{The H$\beta$ (left) and H$\alpha$ (right) profile
decomposition of the variable (top) and constant (bottom)
components. The flux of the constant components is in units
$10^{-15}$\,erg\,cm$^{-2}$\,s$^{-1}$\,\AA$^{-1}$. For the
variable components, the relative increment of the line flux when
the continuum flux increases by an amount of
1$\times10^{-15}$\,erg\,cm$^{-2}$\,s$^{-1}$\,\AA$^{-1}$ is shown.
The correlation coefficient between the variable component of lines
and the continuum
flux at different wavelengths along the line profile is shown in
the middle panels.} \label{profcor}
\end{figure}

As shown by Sergeev et al. (\cite{ser}), if a line profile is
composed of only a variable and a constant part, it is possible
to separate these components. The constant component is present in
the form of narrow emission lines, which are formed far away from
the nucleus and are independent of changes in the continuum
emitted by the central source. The variable component is formed
in the BLR and is strongly dependent on the flux of the central
source through reverberation. We have applied the two-component
profile decomposition method to spectra obtained in 1996 -- 2001.
We assume that the fluxes in the variable part
of the emission lines are correlated linearly with the continuum
flux, as shown by Peterson et al.  (\cite{pet2002}, Fig. 3).
 An underlying continuum was subtracted from each of
the scaled spectra. The region occupied by an emission line is
divided into equidistant narrow spectral intervals (5\AA\ bins).
For every bin a light curve is constructed. A linear regression
for each of these light curves and that of the continuum was
computed. This is done by taking into account the mean delay time
between them ($\sim 20$ days, Peterson et al. (\cite{pet2002})).
The introduction of the more precise time delays
obtained for each year by Peterson et al. (2002) for  H$\beta$
does not
change the profile shape of the variable component, within the
errors marked in Fig.~\ref{profcor}. Since the exact year-averaged
time delay for H$\alpha$ is not known, we used the same
average time delay  for the both lines.  The variable
part in the profile is computed as the increment of line flux per
unit flux increase of the continuum. The non-variable part of the
profile is estimated by the extrapolation of the line flux in
every bin to a zero continuum flux value.

This scheme allows us to separate the part of the line profiles
that is linearly related to the continuum light changes from those
which remain constant in time. It also provides a scheme to
estimate the H$\beta$ and H$\alpha$ narrow line fluxes in an
independent way from that described in Sect.~\ref{sec2.4}. In any
case both methods yielded very similar results.

Fig.~\ref{profcor} shows the variable (top) and constant (bottom)
components of the H$\beta$ (left) and H$\alpha$ (right) lines. The
correlation coefficient of the variable component with the
continuum across the emission lines is plotted in the central panels.
Both variable components present a double-peaked structure with
maxima at radial velocities $\sim\pm 1000$~km/s. The variable
part of the H$\beta$ and H$\alpha$ profiles between $\sim -4000$\,
km/s\, and $\sim +5000$~km/s\, show a highly correlated (r$\sim
0.8$) response to changes in the continuum. The constant
component panels contain mainly the narrow line emission of
H$\beta$, [\ion{O}{iii}]~$\lambda$\,4959, 5007\,\AA\,(left) and
of H$\alpha$, [\ion{N}{ii}]~$\lambda\lambda$\,6548, 6584,
[\ion{S}{ii}]~$\lambda\lambda$\,6717, 6731  and
[\ion{O}{i}]~$\lambda$\,6364 (right).

%%%%%%%%%%%%%%%%%%%%%%%%%%%%%%%%%%%%%%%%%%%%%%%%  fig4

\subsubsection{ Mean and Root-Mean-Square Spectra}
\label{sec3.2.2}

The comparison between an average and root-mean-square (rms)
spectrum provides us a good measure of the profile variability.
Average and rms H$\beta$ and H$\alpha$ profiles were calculated
after removing the continuum and the narrow lines from the
profiles (see Sections~\ref{sec2.3} and~\ref{sec2.4}).

\begin{figure}
\centering
\includegraphics[width=8.8cm]{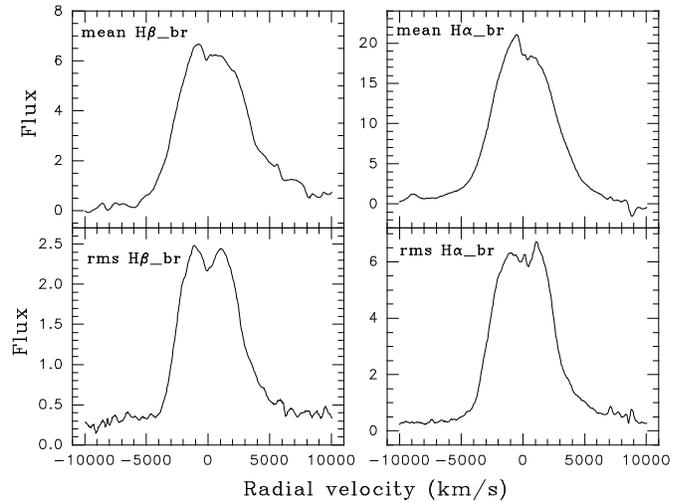}
\caption{
  The mean (top) and rms (bottom) H$\beta$ (left) and H$\alpha$
(right)
profiles. The absissae are the radial velocities relative
to the narrow components; the ordinates give the fluxes in units
$10^{-15}$\,erg\,cm$^{-2}$\,s$^{-1}$\,\AA$^{-1}$.} \label{fig6}
\end{figure}

The mean H$\beta$ and H$\alpha$ profiles and the absolute rms
variations per unit wavelength are shown in Figs.~\ref{fig6}.  It
is clear that both profiles present  a double-peaked structure
with maxima at radial velocities $\sim\pm$1000~km/s relative to the
narrow components. On the mean and rms H$\alpha$  and H$\beta$
profiles, a distinct red asymmetry is observed at radial
velocities  $>2000$\, km/s, the red wing being brighter than the
blue one.  This is indicative of stronger variability of the red
wing as compared to the blue one during the monitoring period.

On the rms H$\alpha$ profile, a small peak is seen at a zero
radial velocity, which is caused by an improper subtraction of the
narrow H$\alpha$ component for spectra of lower resolution. The
FWHM value of the mean and rms profiles is $\sim 6300$~km/s and
$\sim 5800$~km/s, respectively. These values are close to those
obtained by Wandel et al. (\cite{wan1999}) from  spectra obtained
during the 1989 -- 1996. The mean profiles of H$\beta$ and
H$\alpha$ show blue peaks brighter than the red ones during the
1996 -- 2001.

%%%%%%%%%%%%%%%%%%%%%%%%%%%%%%%%    fig5

\subsubsection{  Time evolution of the broad H$\beta$  and H$\alpha$
profiles.} \label{sec3.2.3}

The study of the shape of the broad emission line profiles and
their variation in time, can help in choosing a suitable model
of  the BLR. From our spectra it is seen that within every month,
the broad line profile did not vary at a noticeable level. The
flux in the lines also varied very slightly (as a rule, by 2 --
5\%, and only in some cases up to 10\%).  Within every year, the
shape of the profile did not vary at a noticeable level, but the
fluxes in the broad emission lines varied considerably in
2000--2001. Then averaging the broad profiles over months or
years, in order to increase the  signal to noise ratios,  allows
us to see their  time evolution in a more reliable way than on
the individual spectra. Therefore, for each month and each year,
we have obtained the mean profiles.

\begin{figure}
\centering
\includegraphics[width=8.8cm,bb=84 171 529 690,clip=]{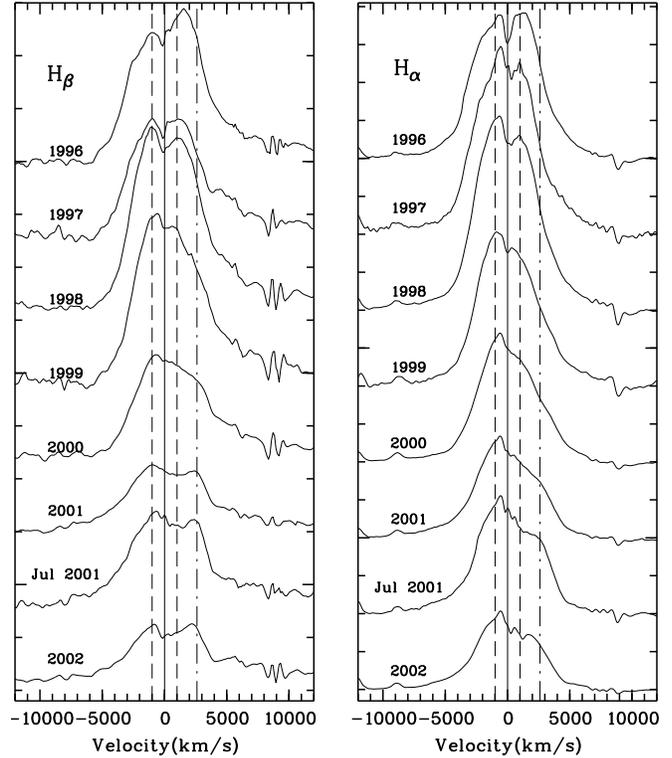}
\caption{ Annual averages from 1996 to 2002 of the observed
profiles of H$\beta$ broad emission component after subtraction of
the continuum and the narrow components of H$\beta$ and
[\ion{O}{iii}]$\lambda\lambda$\,4959, 5007 (left). The same for
H$\alpha$ after subtraction of the continuum and the narrow
components of H$\alpha$, [\ion{N}{ii}]$\lambda\lambda$\,6548, 6584
and [\ion{S}{ii}]$\lambda\lambda$\,6717, 6731 (right). The
vertical lines correspond to the radial velocities: dashed lines:
$\pm 1000$~km/s; dash-dot lines: $\sim +2600$~km/s; thin line:
$0$~km/s. The absissae give the radial velocities relative to the
narrow component of H$\beta$ and H$\alpha$.  The ordinates give
the relative fluxes. The profiles are shifted vertically by a
constant value.} \label{fignew}
\end{figure}

The averaged profiles  of H$\beta$ and H$\alpha$ broad emission
lines in subsequent years are shown in Fig.~\ref{fignew}. The
evolution of the profile is well seen: in 1996, double peaks are
present at radial velocities $\sim -1000$~km/s and $\sim
+1200$~km/s; in  1997 -- 1998, the double peaks are distinctly
seen at radial velocities of about $\pm 1000$~km/s, and in 1999,
they are located at radial velocities of about $\pm
(500-900)$~km/s; in 2000 -- 2002, a bright blue peak is still
present at $\sim$~-(700 -- 1000)~km/s, but on the red side at
$\sim +1000$~km/s, a bending shoulder is seen instead of a bump.
 The relative brightness of the peaks varies: the blue peak was
brighter than the red one in 1998 -- 2002; but in 1996, the red
peak at a radial velocity of $\sim +1200$~km/s became  brighter.
These effects are best seen on the H$\beta$ profile than on the
H$\alpha$ one. It is because the narrow components are not
well subtracted in the H$\alpha$ case.

%%%%%%%%%%%%%%%%%%%%%%%%%%%%%%%%    fig6

In 2000 -- 2002, a new bright bump is clearly seen in the red
wing of the broad lines at a radial velocity of about $\sim
+2500$~km/s. In order to investigate whether or not the
radial velocity of the new bump varies, we compared the observed
profiles of H$\beta$ obtained at different times (before subtraction of the
narrow component). We have measured the
radial velocity of the peak of this bump using good spectra
 with similar resolution (8--9) \AA\, and a high S/N ratio
$\geq$50. We have also used the measurements of the peak
location in May, 2003 obtained from the spectra taken with the 6 m
telescope. The bump is relative broad, and the
determination of the radial velocity of its peak
is somewhat uncertain. Using the
 individual spectra of H$\beta$ we defined the radial
velocities of the bump peak by two methods: 1) as the
mean-weighted location of the point in the bump corresponding to
the level $\sim
0.8$\,Imax  (Imax being the maximum of the peak intensity); 2) by
fitting the bump top with a parabola.
Both methods gave similar results.
The only difference is that the
second method (fitting by a parabola) gives radial velocities of
the bump peak systematically $\sim
100$\,km higher, probably because of the slight blue
asymmetry. We think that the first method gives more realistic results.
Therefore we present the year-averaged radial
velocities of this red peak derived by method 1 in
Table~\ref{tab5n}: 1 --- year; 2 --- the average radial velocity
relative to the narrow component of H$\beta$ and the root-mean-square
error obtained from measurements on
the individual spectra; 3 --- the average continuum flux at
the observed wavelength 5190 \AA\, and its dispersion; 4 --- the S/N
ratio on the average spectrum in the region (5160--5220)\AA.
One sees clearly on this table that the radial
velocity of the peak decreases: in 2000--2001 it
corresponds  to $\sim$ (2500--2600)
km/s, and in 2002--2003 to $\sim$ 2000 km/s, within the uncertainties.
 This effect is well
seen on Fig. 7, where the year-averaged normalized profiles of
H$\beta$, derived by dividing the observed profiles by the
average H$\beta$ flux for each year, are presented. A similar
result is obtained from the observed profiles after subtraction
of the narrow components. Thus, during 3 years (from 2000--2001
to 2002--2003) a considerable shift $\sim  500$ km/s of the red
peak is observed
 on the H$\beta$
profiles.

%%%%%%%%%%%%%%%%%%%%%%%%%%%%%%%%%%%%%%%%%%%%% Table 5

\begin{table}
\begin{center}
\caption{Year-averaged radial velocity of the new red peak in
2000--2003, derived from the observed H$\beta$ profiles before
subtraction of the narrow component} \label{tab5n}
\begin{tabular}{clcc}
\hline \hline
  Year  &  V$_r$ & $F$(cont)* & S/N \\
        &  (km/s)& 5190\AA & (5160-5220)\AA \\
\hline
  2000  &     2660$\pm$106 &    9.27$\pm$1.28 &    70\\
  2001  &     2438$\pm$83  &    9.66$\pm$1.26 &    95\\
  2002  &     2088$\pm$46  &    7.11$\pm$0.98 &    76\\
  2003  &     1986$\pm$75  &    10.43      &    95\\
\hline \multicolumn{4}{l}{* - $F$(cont) in units of
$10^{-15}$(erg\,cm$^{-2}$\,s$^{-1}$\,\AA$^{-1}$)}\\
%%*\rule{0pt}{11pt}  in units of
%%$10^{-15}$\,erg\,s$^{-1}$\,cm$^{-2}$\,\AA$^{-1}$}\\
\end{tabular}
\end{center}
\end{table}

 From Table~\ref{tab5n}, there is no apparent correlation
  between the variations of the radial velocities of this peak and
the average continuum flux.

%%%%%%%%%%%%%%%%%%%%%%%%%%%%%%%%    fig7
\begin{figure}
\centering
\includegraphics[width=8.8cm,bb=60 60 555 745,clip=]{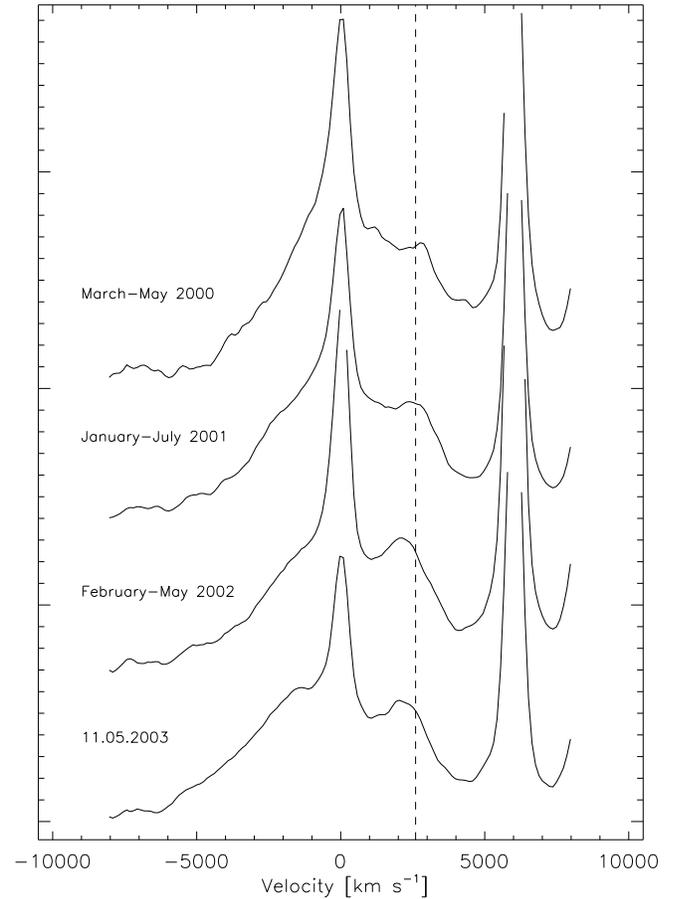}
\caption{ Year-averaged normalized profiles of H$\beta$ in
2000-- 2003. The vertical dashed line corresponds to a radial
velocity  $ +2600$~km/s. The absissae give the radial
velocities relative to the narrow component of H$\beta$. The
profiles are shifted vertically by a constant value.}
\label{fig_sub_hb}
\end{figure}

%%%%%%%%%%%%%%%%%%%%%%%%%%%%%%%%%%%%%%%%%%%%%%%%%%%%%%%%%%%FIG7,8

\begin{table*}
\begin{center}
\caption{ Year-averaged radial velocity of the blue and red bumps
derived from the difference H$\beta$ and H$\alpha$ profiles}
\label{tab8}
\begin{tabular}{lcccccc}
\hline \hline Year&    \multicolumn{3}{c}{H$\beta$ radial velocity
(km/s)}&
\multicolumn{3}{c}{H$\alpha$ radial velocity (km/s)}\\
&  $V_{b}$&   $V_{r1}$&   $V_{r2}$&  $V_{b1}$&   $V_{r1}$&  $V_{r2}$ \\
\hline \hline 1996& $-906\pm 131$ & $+1461\pm ~56$& &   $-1200\pm
~69$
& $+1379\pm 103$&       \\
1997& $-833\pm 185$ & $+1216\pm 164$&                &   $
~-696\pm ~32$
& $+1079\pm ~64$&       \\
1998& $-994\pm ~91$ & $+1137\pm 103$&                & $-1075\pm
~67$
& $+1026\pm ~71$&       \\
1999& $-704\pm 200$ & $ ~+838\pm 155$&                & $~-944\pm
135 $&
$~+764\pm 119$&        \\
2000& $-544\pm 134$ & $ +817:  $&  $+2863\pm 262$& $ ~-775\pm 192
$&
$+800:   $&  $+2787\pm 261$\\
2001& $-704\pm 316$ & $+954:  $&  $+2707\pm ~70$&  $ -1043\pm 172$
& $ +800:  $&  $+2526\pm ~92$\\
2002& $-938\pm 394$ & $+674: $&  $+1898\pm 200$&  $ -1158\pm 144$
& $ +584:  $&  $+1708\pm 191$\\
\hline
\end{tabular}
\end{center}
\end{table*}

\subsubsection{ H$\alpha$ and H$\beta$ difference profiles}
\label{sec3.2.4}

%%%%%%%%%%%%%%%%%%%%%%%%%%%%%%%%    fig8
\begin{figure}
\centering
\includegraphics[width=8.5cm,bb=84 183 519 687]{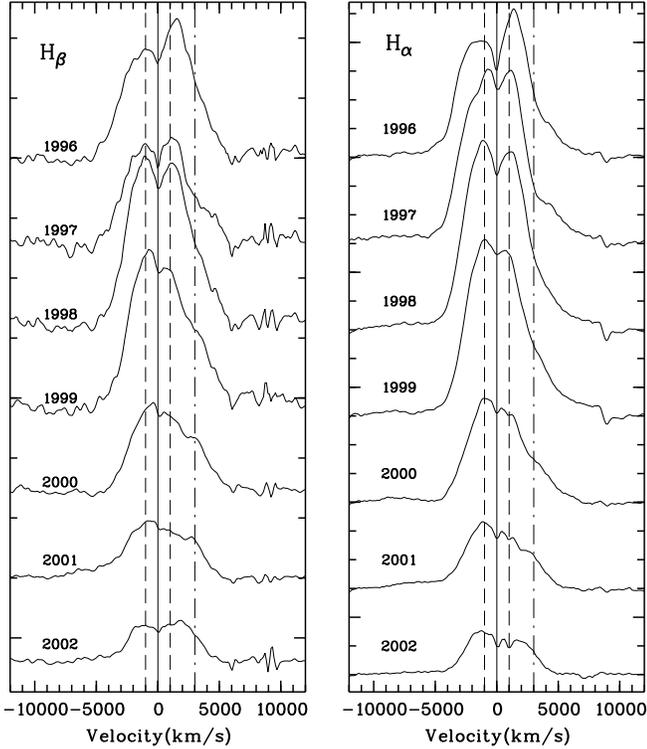}
\caption{Annual averages from 1996 to 2002 of the difference
profiles of the broad H$\beta$ (left panel) and H$\alpha$ (right
panel) lines. The vertical lines correspond to the radial velocities:
 dashed lines: $\pm 1000$~km/s;  dash-dotted lines:
 $\sim +3000$~km/s; thin lines:
$0$~km/s. The absissae represent the radial velocities relative
to the narrow component of H$\beta$ and H$\alpha$.  The ordinates
give the relative fluxes. The profiles are shifted vertically by
a constant value. } \label{fig17}
\end{figure}
%%%%%%%%%%%%%%%%%%%%%%%%%%%%%%%%%%%%%%%%%%%%%%%%%%%%%%%%%%%%%%%%%%%%%%%%%%%%%%

For the analysis of the shape changes in the broad line profiles,
the best method is that of direct subtraction of spectra, i.e.
obtaining the difference spectra. In this case, the narrow
emission lines and the absorption lines of the host galaxy, which
may distort the observed emission line profile, are fully
cancelled. In order to carry out a proper subtraction, it is
necessary that the spectra had comparable spectral resolutions. To
obtain difference spectra, one usually subtracts either the mean
spectrum or a spectrum representative of the minimum activity
state. We have adopted the second alternative, as the spectrum in
the minimum state, provides us with crucial information about the
contribution of the narrow emission and absorption lines, which
are to be removed in order to carry a proper study of the profile
shape changes with time. Hence, we obtained the H$\beta$ and
H$\alpha$ difference profiles by subtraction of the minimum
activity state spectrum, represented by the mean spectrum of the
May 15, 17 and June 4, 2002 observations for H$\beta$, and June
4, 2002 for H$\alpha$. The subtraction was performed by our
scaling program (see Appendix A), and a spectrum in minimum
state  was used as a reference. Since within every year, the
shape of the line profiles changes only slightly, we have
calculated the annual mean difference profiles for H$\beta$ and
H$\alpha$ (Fig.~\ref{fig17}).  Variations of the profiles similar
to those on the observed H$\beta$ and H$\alpha$ line profiles (see
section~\ref{sec3.2.3}) are well seen. Peculiarities in the shape
of the bumps in the difference profiles are better seen than on
the observed profiles. In the H$\alpha$ difference profiles for
2000--2002, at radial velocity $\sim +1000$~km/s a drop in
brightness (dip) is seen, this is due to over-subtraction of the
[NII]~$\lambda$6584 line. This due to the fact that the FWHM of
this line is smaller than the one for the [OIII]~$\lambda$5007
line, used to reduce the spectra to similar resolution (see
Section 2.4). This effect becomes noticeably on the 2000--2002
spectra, when the flux in the broad components decreases by
factors of the order of 2 to 3. From the monthly-averaged
difference profiles of H$\beta$ and H$\alpha$, we defined the
radial velocities of the peaks as the mean weighted location of
the bump tops (for details see Section 3.2.3). Our results are
listed in Table~\ref{tab8}, there the annual averages for the
radial velocities of the blue ($V_b$), and red ($V_{r1}$ and
$V_{r2}$) peaks along with their corresponding dispersion values
are presented, for H$\beta$ and H$\alpha$. In Table~\ref{tab8},
it is seen that the peak radial velocities of H$\beta$ and
H$\alpha$ for different years agree with each other within the
errors. It may be noted that during 1996--1999, the radial
velocities of the blue ($V_{b}$) and red bumps ($V_{r1}$) were,
on average, close to $\pm 1000$~km/s. However, it is distinctly
seen that in 1996 the radial velocity of the red peak ($V_{r1}$)
was larger than in subsequent years. During 2000 -- 2001, a
decrease of the radial velocity of the blue ($V_b$) peak in the
H$\beta$ difference profiles was observed. At that time the
radial velocity determinations of the red peak ($V_{r1}$) from
the difference profiles in 2001--2002 are uncertain (marked with
a colon in Table~\ref{tab8}), because in this velocity zone, a
bending shoulder is present in the H$\beta$ profile, and the
previously mentioned dip in the H$\alpha$ profile, due to
over-subtraction of [NII]~$\lambda$6584, affects our results. In
2000 a new distinct red bump appeared with a radial velocity of
about $+2800$~km/s ($V_{r2}$ in Table~\ref{tab8}). By 2002 its
radial velocity had decreased to about $+1800$~km/s, while its
brightness became similar to that of the blue bump. We can
definitely say that in the year 2000 a new bump in the red wing
of the lines appeared and that its radial velocity decreased by
about 1000~km/s between 2000 and 2002.

\subsection{ Flux variability for different parts of
H$\alpha$ and H$\beta$ profiles} \label{sec3.3}

\begin{figure}
\includegraphics[width=8.5cm]{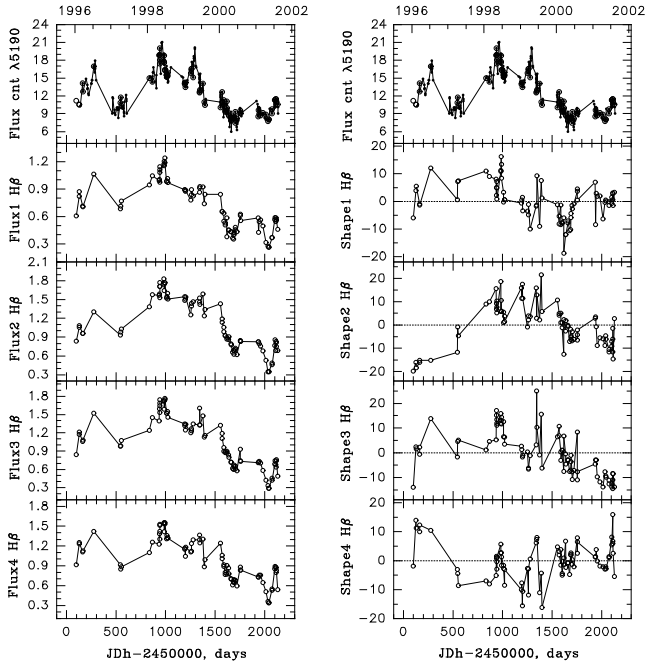}
\caption {Variations of the flux (left panel) and shapes (right
panel) in different parts of the H$\beta$ profile, and in the
``combined" continuum at the observed wavelength $\lambda
5190$\AA. The radial velocity intervals of the profile segments:
%are given in Section~\ref{sec3.3}
the H$\beta$1, H$\alpha$1 set goes from $-3000$~km/s to
$-2000$~km/s; the H$\beta$2, H$\alpha$2 set goes from $-1500$~km/s
to $-500$~km/s; the H$\beta$3, H$\alpha$3 set goes from $+500$~km/s
to $+1500$~km/s ; the H$\beta$4, H$\alpha$4 set goes from
$+2000$~km/s to $+3000$~km/s. The flux in the lines are given in units
of $10^{-13}$\,erg\,cm$^{-2}$\,s$^{-1}$, and in the continuum
are given in units of
$10^{-15}$\,erg\,cm$^{-2}$\,s$^{-1}$\,\AA$^{-1}$. The shapes are
in relative units (see Section~\ref{sec3.3.2}).} \label{fig10}
\end{figure}

\begin{figure}
\includegraphics[width=8.5cm]{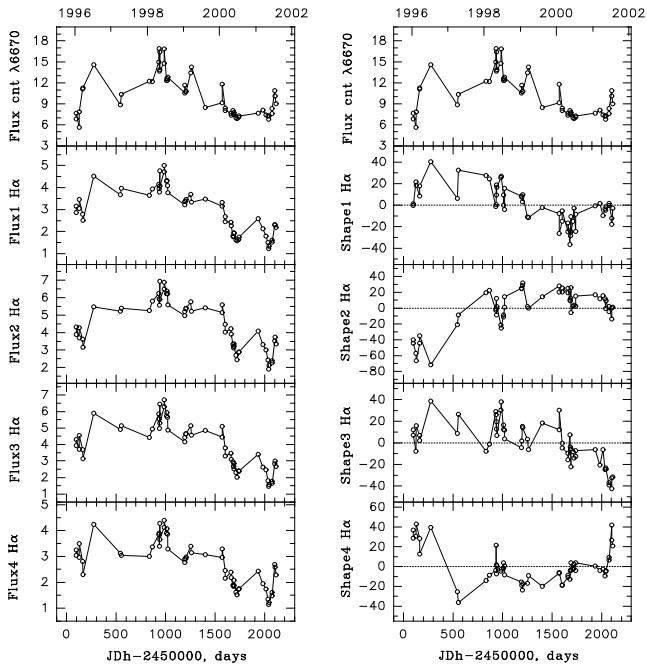}
\caption{
Same as Fig.~\ref{fig10}, but for H$\alpha$ and for
the continuum at $\lambda 6670$.}
%Figures 10,11 are only available in the electronic form
\label{fig11}
\end{figure}

As mentioned
above, the flux of the H$\beta$ and H$\alpha$ emission lines as
well as the $\lambda\lambda$5190\ continuum flux, varied
significantly between 1996 and 2002. Some parameters of the
variability were noted in Section~\ref{sec3.1}. However, as it is
shown in Sections \ref{sec3.2.3},~\ref{sec3.2.4}, important
details such as bumps or oblique shoulders are present in the
profiles of the broad lines. Therefore, it is of great interest
to study the behaviour in time of these parts of the profiles
relative to each other, and with respect to continuum
variations. To this purpose, we have divided the emission line
profiles into several radial velocity bins and constructed the
data sets for a number of  time series that are of interest. We have
chosen symmetrical bins relative to zero radial velocity of the
H$\beta$ and H$\alpha$ line profiles. These bins include either
distinct peaks or notable features that were observed at
different times (see Sections~\ref{sec3.2.3},~\ref{sec3.2.4}):
the H$\beta$1, H$\alpha$1 set from $-3000$~km/s to $-2000$~km/s;
the H$\beta$2, H$\alpha$2 set from $-1500$~km/s to $-500$~km/s;
the H$\beta$3, H$\alpha$3 set from $+500$~km/s to $+1500$~km/s ;
the H$\beta$4, H$\alpha$4 set from $+2000$~km/s to $+3000$~km/s.
The flux associated to those radial velocity bins are named as:
Flux1, Flux2, Flux3, Flux4, respectively. Their light curves for
H$\beta$ and H$\alpha$ are plotted in
Figs.~\ref{fig10},\ref{fig11} (left).

The upper left and right panels present the light curves for the continuum near
H$\beta$ (Fig.~\ref{fig10}) and H$\alpha$ (Fig.~\ref{fig11}). From these figures,
one can  see that the flux in different parts of  H$\beta$ and H$\alpha$
change in an almost identical manner.

%%%%%%%%%%%%%%%%%%%%%%%%%%%%%%%%    fig9

%%%%%%%%%%%%%%%%%%%%%%%%%%%%%%%%    fig10

\subsubsection{The shapes of the H$\alpha$ and H$\beta$
in relation to
their variability} \label{sec3.3.2}

The broad H$\beta$ and H$\alpha$ emission lines show changes in both: flux and
shape of the profiles. Wanders \& Peterson (\cite{wan1996}) suggested a method
for studying the variations of the profile shape by means of a normalization of the
emission lines by their total flux and then see how these normalized profiles
change in the course of time. They named shape the function $Fq(v,t)$,
defined as:

\begin{equation}
Fq(v,t)=F(v,t)- [<F(v,t)>/<F(t)>]\cdot F(t),
\label{equ5}
\end{equation}

where  $F(v,t)$ is the flux in the profile at a radial velocity
$v$ at a time $t$; $<F(v,t)>$ and $<F(t)>$ are the time-averaged
 (over the whole set of observations) flux at a radial velocity
$v$ and the total flux in the line, respectively; $F(t)$ is the
total flux in the emission line at a time $t$. By definition one gets
%, the integral from
$\int$Fq(v,t)dv=0, if the integration is performed over the whole emission
line. The shape value in a given radial velocity interval (or the
shape value in a given profile part) is defined by
integrating equation~\ref{equ5} in this velocity interval. From
equation~\ref{equ5}, it follows that the value of $Fq(v,t)$ will be
close to zero when the profile part $ F(v,t)$ changes
proportionally to the total flux $ F(t)$. If the proportionality
is violated due to delay effects or to some other
reason, then $Fq(v,t)$ will be either positive or negative. A
positive value $Fq(v,t)$ means that  at a time $t$ the given profile
 part is more prominent than in the average
profile, while a negative value means that this part is less
prominent. So, $F(v,t)$ is a
sensitive indicator of the relative prominence of a part of the
profile.
If the profile shapes vary due to reverberation effects,
$Fq(v,t)$ must be correlated with the
flux variations of the continuum after some delay.

We have considered four shape time series in the radial
velocity intervals mentioned in section~\ref{sec3.3}:
 shape1 for ($-3000$, $-2000$)~km/s;
 shape2 for ($-1500$, $-500$)~km/s;
 shape3 for ($+500$, $+1500$)~km/s and
 shape4 for ($+2000$, $+3000$)~km/s).

The changes of the shape of the different profile parts are shown
on the right panels of Figs.~\ref{fig10} and \ref{fig11}, for
H$\beta$ and H$\alpha$, respectively. The changes in shape behave
differently from continuum variations: sometimes they change
following the continuum changes, and some other time in quite the
opposite manner. The behavior of the shape of different parts of
the lines differs, though the changes of a given shape for H$\alpha$ and
H$\beta$ are similar.

\subsubsection{The relations flux-shape and shape-continuum for
the different parts of the H$\beta$ and H$\alpha$ broad emission
lines} \label{sec3.3.3}

\begin{figure}
\resizebox{\hsize}{!}{\includegraphics{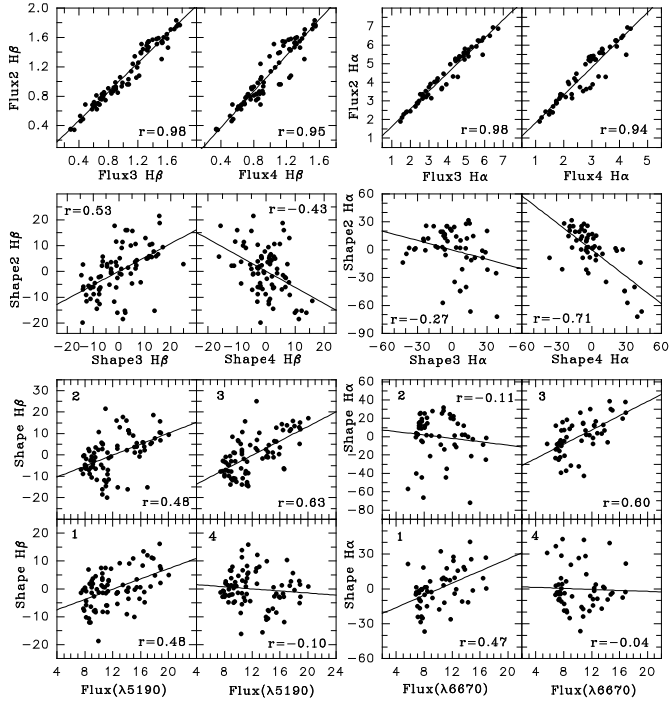}}
\caption{Flux-flux, shape-shape and continuum flux -shape
correlations for different parts of the H$\alpha$ and the
H$\beta$ lines (the number shape is inside the plot). The
correlation coefficients $r$ are given inside the plot. }
\label{fig11d}
\end{figure}

A clear-cut distinction in the behavior of the flux and the shapes
can be seen
 in Fig.~\ref{fig11d}, where we show the flux-flux, shape-shape and
shape-continuum flux correlations for the different parts of the emission
profiles of  H$\beta$ and H$\alpha$.

%%%%%%%%%%%%%%%%%%%%%%%%%%%%%%%%    fig12

There is a high degree of correlation between the flux in
the various parts of H$\beta$ and H$\alpha$
(Fig.~\ref{fig11d}, top)). The correlation coefficient values
being 0.94 -- 0.98. Yet, the correlations for the shapes of
various parts of the lines are quite different. For example,

\begin{itemize}

\item The shape4 and shape2 have a tendency to be anti-correlated with
each other.

\item Shape4 in H$\beta$\ and H$\alpha$, including the bump at
radial velocity +2500~km/s, does not correlate with the continuum.

\item Shape2 in H$\beta$\ is weakly related to the continuum, but the
shape2 in H$\alpha$\ is not.

\item Shape3 and shape2 have a tendency to be correlated in
H$\beta$ and weakly anti-correlated in H$\alpha$.

\item A good correlation with the continuum for the shape3 in
 H$\beta$ and H$\alpha$ is observed.
\end{itemize}

Note that the analysis of shape2 and shape3 in H$\alpha$, should
be done  with care, because the narrow components of H$\alpha$ and
nitrogen are not properly subtracted in some profiles. Hence,
the derived fluxes are subject to uncertainties, and the profiles
in these regions may be considerably distorted. In the H$\beta$
case, the narrow component is well subtracted and data are more
reliable. Therefore, from the H$\beta$ analysis alone, for shape2
and shape3 that include the double-peaks at radial velocities
close to $\pm1000$~km/s, we conclude that they are weakly
correlated with each other and also with the continuum.

Thus, the shape of the individual parts of the line profiles changes,
unlike their fluxes, in a more complicated way. So, since the changes
in shapes correlate sometimes slightly with the continuum changes, or do not
correlation with them, then the changes may not be related to reverberation
effects, and some other mechanisms should be invoked for their explanation.

\subsection{ The Balmer Decrement}
\label{sec3.4}
\subsubsection{Integral Balmer Decrement}
\label{sec3.4.1}

\begin{figure}
\centering
\includegraphics[width=8.5cm]{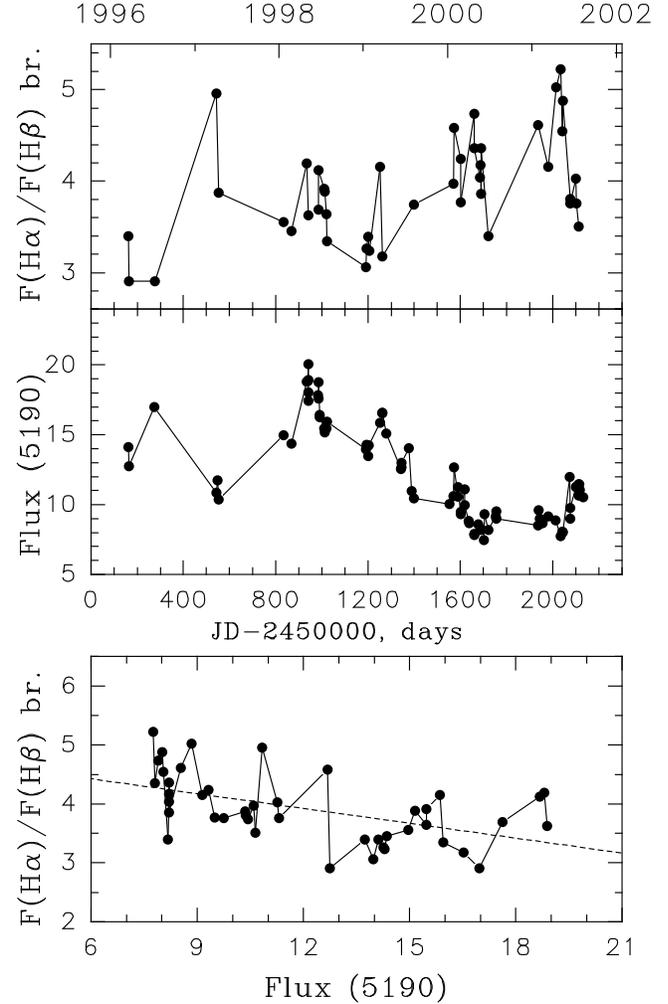}
\caption{Variations of the Balmer decrement $F(H\alpha)/F(H\beta)$
(top) and of the continuum flux (middle) in 1996 -- 2001;
variations of the Balmer decrement versus those of the continuum
flux (bottom).} \label{fig18}
\end{figure}

>From the spectra with the subtracted narrow component template
(see section~\ref{sec2.4}), we determined the fluxes of H$\alpha$
and H$\beta$ emission line broad components within the radial
velocity intervals $\pm 6000$~km/s. The integral Balmer decrement
(BD) is the flux ratio $F(H\alpha)$/F$(H\beta)$. Note that
the FeII (42 multiplet) line  at the rest wavelength 4924 \AA\
corresponds
 to a radial velocity of +3874 km/s relative to the
narrow component of H$\beta$. This line sets in the
broad red wing of H$\beta$ close to the blue
wing of the emission line [OIII] 4959. No measurable feature is
seen on our spectra at this place. Since it is known that the
intensities of FeII42 (4924, 5018, 5169) lines are comparable,
 we have estimated the contribution of FeII42
5018 \AA\ line. This line is well seen on our
spectra. We have obtained for the ratio of
FeII42(5018) to the flux integrated within the interval of
radial velocities +-6000 km/s (i.e. onto the broad component of H$\beta$):
 F(FeII42)/F(H$\beta$) $\sim
(0.02-0.03)$ in 1996--1999, and $\sim (0.03-0.06)$
 in 2000--2001.
Thus, the contribution of FeII42 to the H$\beta$
broad component is  $\sim (2-3)$\% in 1996--1999,
and $\sim (3-6)$\% in 2000--2001. This is within the measurement
errors of the Balmer decrement. Therefore, we did not take into
account the contribution of FeII42.
Fig.~\ref{fig18} shows the
behavior of Balmer decrement (upper panel),the $\lambda$ 5190
continuum, (middle panel), and the relation between changes of
the Balmer decrement and continuum flux (bottom panel). There an
anticorrelation between the changes of the Balmer decrement and
the continuum flux is evident (correlation coefficient $\sim 0.52
$, bottom panel in Fig.~\ref{fig18}).

%%%%%%%%%%%%%%%%%%%%%%%%%%%%%%%%    fig13

%%%%%%%%%%%%%%%%%%%%%%%%%%%%%%%%    fig14
\begin{figure}
\centering
\includegraphics[width=8.8cm,bb=129 201 472 655]{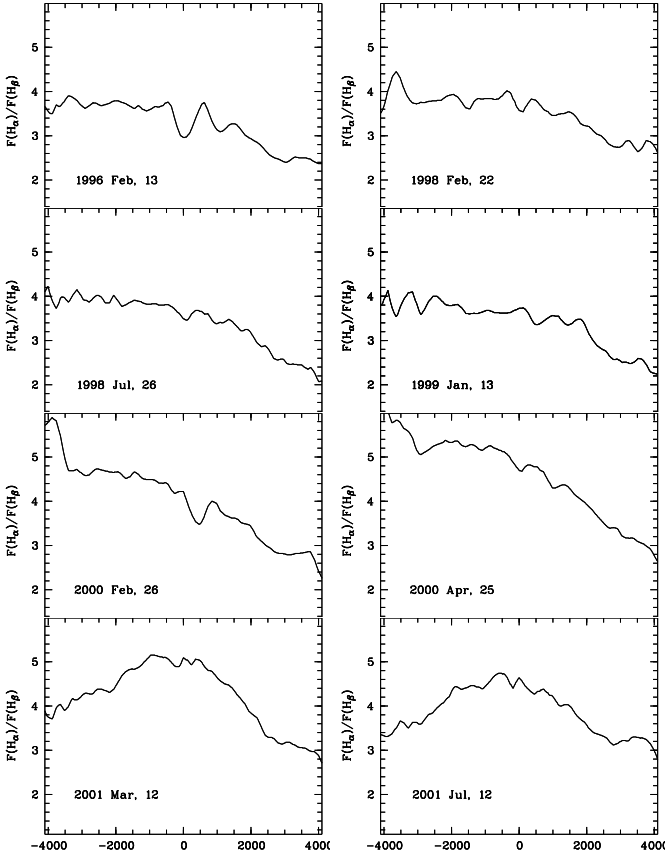}
\caption{ Variations of the Balmer decrement $F({\rm H}\alpha)/F({\rm
H}\beta)$  along some individual profiles of the broad
lines. The absissae represents the radial velocities relative to
the narrow components.} \label{fig19}
\end{figure}

The same tendency is present in the annual averaged values of the BD
and the continuum fluxes. In Table~\ref{tab9}, the mean values of
the Balmer decrement are listed
 for several years together with the mean values of the continuum flux.

\begin{table}
\caption{The anuual averaged Balmer decrement and continuum fluxes
$F$(cont) at the observed wavelength $\lambda 5190$\,\AA }
\label{tab9}
\begin{tabular}{ccc}
\hline
\hline
Year&     $F$(H$\alpha$)/$F$(H$\beta$)&  $F$(cont)* \\
\hline
1996 &     3.48$\pm$ 0.45 &       14.62$\pm$  2.2 \\
1998 &     3.51$\pm$ 0.44 &       16.96$\pm$  1.69\\
1999 &     3.51$\pm$ 0.44 &       14.05$\pm$  1.62\\
2000 &     4.22$\pm$ 0.42 &        9.27$\pm$  1.28\\
2001 &     4.33$\pm$ 0.58 &        9.66$\pm$ 1.26\\
\hline \multicolumn{3}{l}{* - $F$(cont) in units of
$10^{-15}$(erg\,cm$^{-2}$\,s$^{-1}$\,\AA$^{-1}$)}\\
\end{tabular}
\end{table}

>From Table~\ref{tab9} it is seen that during 1996 -- 1999 the
average flux in the continuum varied slightly, while the Balmer
decrement did not changed. However, in 2000 -- 2001 the average
flux in the continuum decreased and the Balmer decrement became
considerably larger (steeper).

\subsubsection{ Balmer Decrement across some individual profiles}
\label{sec3.4.2}

Fig.~\ref{fig19} shows the Balmer decrement
across some individual broad lines profiles.
The decrement is similar to that of other profiles
for the same years.  Yet, the Balmer decrement across
the line profile changes from year to year. During 1996 -- 1999
in the radial velocity interval from $\sim -4000$~km/s to $\sim
+2000$~km/s, the Balmer decrement varied only slightly and had a value
of $\sim 4.0$, while in the radial velocity interval
$+(2000-4000)$~km/s, it decreased, from $\sim 4.0$ to 2.0. In
the year 2000, the Balmer decrement increased, as a whole. Yet, a monotonic decrease
with radial velocity is clearly present. During 2001, maximum values ($\sim 5$) were observed
at the line center, and at the edges the BD decreases with a
somewhat larger gradient in the red wing.

\section{Summary of our main results and comparison with other results
from the literature}
\label{sec4}

\begin{enumerate}
\item

 In Section~\ref{sec3.2} we have shown that
the observed mean and rms profiles as well as the annual averaged,
observed and difference profiles
of H$\alpha$ and H$\beta$ present a double-peak structure, i.e.
bumps at radial velocities $\sim\pm1000$~km/s ( zero velocity
being defined as that of the narrow components). While the double
peaks were obtained after subtraction of the narrow line
components, a question arises about the reality of these features,
as it dependent on the method of subtraction of those components.
The following facts are taken as evidence to the reality of the double
peaks: 1) Double-peaks are obtained on the variable component
when the original H$\beta$ and H$\alpha$ profiles are decomposed,
before subtraction of the narrow components (Section
~\ref{sec3.2.1}); 2) The appearance of double peaks on the
difference profiles of H$\beta$ and H$\alpha$, when the narrow
lines are subtracted automatically (Section~\ref{sec3.2.4}).

The relative intensity of these peaks varied: in 1996, the red
peak was brighter than the blue one, and viceversa, in
1998--2002, the blue peak became brighter. We have seen that the
radial velocities of the double peak vary within $\sim\pm$(500 --
1200)~km/s.

The double-peaked structure in the central part of the broad
lines in the radial velocity range $\sim\pm$ 500 -- 1500~km/s was
observed earlier. Double-peaks at velocities $\sim\pm500$~km/s
were seen in the annual averaged profiles of H$\beta$ in 1986 --
1987 by Wanders and Peterson (\cite{wan1996}). On the individual
difference profiles of H$\beta$, the double peaks were also
evident in 1992 (Iijima \& Rafanelli~\cite{iij}). Very distinct
double peaks were also noticeable on the difference H$\alpha$
profiles in July 1986 - June 1985 (Stirpe et al. ~\cite{sti1988})
and in May 1987 - July 1986 (Stirpe \& de Bruyn ~\cite{sti1991}).

\item

During 2000 -- 2002, when the flux of the continuum and the lines
greatly decreased, a new distinct peak  at a radial velocity of
$\sim +2500$~km/s appeared on the profiles of H$\alpha$ and
H$\beta$. This peak had comparable brightness to the blue peak at
a radial velocity of $\lesssim -1000$~km/s)(Section
~\ref{sec3.2.3}). A radial velocity decrease of $\sim 500$~km/s
for this feature occured between 2000 and 2002 (Table~\ref{tab5n}
and Fig.~\ref{fig_sub_hb}). The same peak is seen on the
difference profiles of H$\alpha$ and H$\beta$ (Section
~\ref{sec3.2.4}). Its radial velocity decreased as well, but by a
larger value ($\sim 1000$~km/s) than the one derived from the
observed profiles. This is a result of the details present in the
line profiles at minimum activity state, which are subtracted
from the observed profiles. In the red wing of H$\beta$ at
minimum activity state, a distinct wide asymmetric weak bump is
observed at a radial velocity of about $+2200$~km/s. Therefore on
the difference profiles in the red wing, we observe excess
emission relative to this weak bump in the spectrum at minimum
state. Thus, from our data set, we can establish the appearance
of a new peak in the red wings of H$\alpha$ and H$\beta$ during
2000 -- 2002. The radial velocity of this peak in 2002 decreased
by $\sim$(500 -- 1000)~km/s, with respect to the value determined
for the year 2000.

\item
>From the  decomposition  of H$\alpha$ and H$\beta$ profiles (see
Section ~\ref{sec3.2.1}) we have found that the fluxes of the
variable component of the lines correlate well with the flux
variations of the continuum. In the radial velocity interval
$\sim -4000$~km/s to $\sim +5000$~km/s the correlation
coefficient has a high value r~$\sim$(0.8 -- 0.9). In
Section~\ref{sec3.3}, we have shown that the flux of different
parts of the profiles, change in an almost identical manner being
highly correlated both with each other (r~$\sim$ 0.94 -- 0.98)
and with the continuum flux levels (r~$\sim$ 0.88 -- 0.97). These
facts indicate that the flux variability in different parts of
the line profiles on short time scales is caused mainly by the
reverberation effect.

\item

Line profile shapes evolve slowly, on time scales of months
to years. The shape variations of the different parts of the line
profiles, mildly correlate with each other and/or with the continuum
variations, or simply do not correlate at all (see Sec.~\ref{sec3.3.3}
and Fig.~\ref{fig11d}). So, the profile shape changes on long
time scales are not due to reverberation, as earlier discussed
by Wanders \& Peterson (\cite{wan1996}).

\item

 The flux ratio of $F$(H$\alpha)/F$(H$\beta)$ (BD) shows a
clear anti-correlation with the continuum flux (see
Fig.~\ref{fig18} and Table~\ref{tab9}). Such an effect has also
been observed in other monitoring campaigns of Sy galaxies
(Dietrich et al.~\cite{die1993}; Shuder~\cite{shu}). In addition,
there are changes of the Balmer decrement across the profiles and
it varies from year to year (see Fig.~\ref{fig19}).
\end{enumerate}

\section{Discussion}
\label{sec5}

The study of the H$\alpha$ and H$\beta$ broad line profiles in
NGC 5548 with different techniques, shows the presence of
double-peak structures at radial velocities
$\sim\pm1000$~km/s relative to the narrow steady component (Section
~\ref{sec4}, point 1).

The appearance of double peaked structures in the profiles of
broad emission lines is predicted by some types of models: a)
different versions of accretion-disc models (Dumont \&
Collin-Souffrin ~\cite{ca,cb}, Rokaki et al. ~\cite{rok2}, Chen \&
Halpern 1989, Eracleous et al.~\cite{era}); b) bi-conic gas flows
(Veilleux and Zneng~\cite{vei}); c) binary black holes (Gaskell
~\cite{gas1983}).

The strongest argument against the binary black hole model is
the fact that the velocity curve of the H$\alpha$ red peak in
3C390.3 (the most prominent object with double peaks) is
inconsistent with the best-fitting binary models (Eracleous et
al., ~\cite{era1997}). Within the framework of the model of
binary black hole, the appearance of the third peak  in NGC 5548
at a radial velocity of $\sim +2500$ km/s is impossible to
explain. The biconical gas flows imply mainly radial motions in
BLR, not supported by any observations. For instance, from the
CCF analysis of the H$\beta$ emission line profiles from the data
of AGN-Watch monitoring in 1989--1993, Wanders and Peterson
(\cite{wan1996}) excluded the possibility of large radial
motions in BLR of NGC 5548. Kollatschny and Dietrich (\cite{kol})
showed that the outer wings of the Balmer lines have a
tendency to repond faster than their
cores to the continuum variations. We also tried
to carry out the CCF analysis for
different parts of the H$\alpha$ and H$\beta$ profiles using
our data. But because of our poor sample there were large
uncertainties in the time delay determination. However, in spite of
this, we noted that H$\alpha$ and H$\beta$ show a same
tendency for smaller time delays at high velocities
delays than at low velocities, and that there is no time delay
 between the blue and red wings of
H$\beta$ within 3--4
days. Of course, our results are only qualitative,
 but they do not contradict the
results of the AGN-Watch monitoring. Thus, the available observations
indicate predominantly rotational (or virial) motions of
NGC 5548 in BLR, and support the scenario of formation of the
broad emission lines in an accretion disk.  Assuming a standard
geometrically thin disc, Dumont and
Collin-Souffrin~(\cite{ca,cb}) have  shown that the broad Balmer
lines can be almost entirely due to disc emission. They obtained
a large variety of profiles, including double peak symmetrical
profiles for non-relativistic cases. Relativistic corrections
introduce profile asymmetries: the blue side of the profile
becomes brighter than the red one. However, in 1996, the red peak
was brighter than the blue one, while at other times
(1998--2002), the blue peak was the brighter one. The cases when
the red peak is brighter than the blue one, are in contradiction
with the predictions of relativistic circular disc models.
However, as the calculations of Eracleous et al.~(\cite{era})
show, such case is possible, when the broad lines originate in a
relativistic, eccentric disc. In this case the changes in the
line profile shapes, will be mainly due to precession of the
disc. If this is the case, the radial velocities and the
intensities of the double peaks would change slowly (in
timescales of months to years), without following the effects of
reverberation.

An alternative explanation for the observed intensity variations of
the double peaks relative to each other, is to involve local
disc inhomogeneities, which may be responsible for some of
the substructures of the emission line profiles. Actually it is
now clear that at the  BLR distance, the disc becomes
gravitationally unstable (Collin \&
Hur\'e~\cite{1999A&A...341..385C}). At that point the disc is expected to be
highly inhomogeneous, including clouds and possibly spiral arms
and it is certainly not constituted by a uniform density medium
(e.g. Bunk, Livio \& Verbunt~\cite{bun}; Chakrabarti \&
Wiita~\cite{cha}).

Another issue which must be taken into consideration, is the fact
that the line emitting regions should be illuminated by the
central continuum source, in order to be able to reprocess an important
fraction of such radiation.
It is now well established that the UV-X source has  small
dimensions as compared to the BLR (say 10-100$R_G$, where $R_G$ is the
 gravitational radius,
  the BLR being located at 10 $^{3-4} R_G$). So, if the
 BLR constitutes itself the outskirts of the disc, it should be
 {\it able to see} the central source. There to this aim, there are several
 possibilities:

\begin{itemize}

\item  As suggested by Dumont \& Collin (1990a, 1990b), the central radiation
can be back-scattered towards the disc, as most of the hard
radiation is emitted in the soft X-ray range ($\ll$ 10 keV). Yet, this
requires a highly ionized Thomson thick medium, surrounding the
disc and the central source. This medium would create a
low energy cut-off (at many keV), which is not
observed in the high energy spectrum of AGNs and would change dramatically
the hard
X-range spectrum due to multiple Compton scattering.

\item The disc should  extend vertically, up to a scale height equal to
at least one tenth of the radius. This would imply that a
good
fraction of the observed velocity (tens of percent) shall be
non-rotational in origin. Probably due to large scale chaotic gas dynamical
motions --macro-turbulence-- added on top of regular rotation.

 \item The disc can be geometrically thin but warped, as suggested by recent
 observations
(Kinney et al. 2000). The warping could be due to self-irradiation
instabilities (Pringle 1996), to misalignment of the spin of the BH with the disc
 (Bardeen \&
Petterson 1975), or to disc feeding through a
misaligned inflow.

\end{itemize}

In 2000 -- 2002, we detected a new peak at a radial
velocity of
 $\sim+2500$~km/s in the red wing of  H$\alpha$ and H$\beta$.
This bump moved gradually along the line profile and its radial
velocity decreased between  2000 and  2002 by $\sim 500$~km/s on
the observed profiles and by  $\sim 1000$~km/s on the
differential profiles (Section ~\ref{sec4}).
 On the normalized
H$\beta$ profiles (1989 -- 1993), Wanders and Peterson
(\cite{wan1996}) noted the periodical appearance of broad red- and
blue-wing shoulders at radial velocities of $\sim \pm
2500$\,km/s. According to these authors,
their relative intensities  change over the years, but not their positions
in the line profile. They interpreted this fact with a model of
cloud illumination by an anisotropic source. However, on the
 normalized H$\beta$ profiles shown by these authors, no
distinct peaks (bumps) are noted. It looks more as an asymmetry in
the line profiles, i.e. a brightness increase within the velocity
interval $\pm 1000-2500$\,km/s. It was impossible to determine
the location of the peaks from these profiles, since they were
absent. Possibly, during the period of 1989--1993, the components
appearing  periodically  in the velocity interval $\pm
(1000-2500)$\, km/s  were very broad and of low contrast.
Therefore, the averaged profile obtained after  summation of a
large amount of spectra, showed the brightness increase in the
form of inclined shoulders. Note that in 1991--1993 Sergeev et
al. (\cite{ser}) observed a distinct low contrast broad emission
feature in the far red wing of the H$\alpha$ profile. The radial
velocity of this feature decreased from $\sim +4500$\,km/s in
1991 to $\sim +2100$ km/s in 1993. So, from our observations and
those of Sergeev et al. (\cite{ser}), we deduce that the position
of the bump changes  with time in
 the line profile, and therefore the
interpretation of Wanders and Peterson (\cite{wan1996}) based on
a constant radial velocity of the bump is not correct.
We think that the appearance of such
details (bumps) moving along the profile is due to
inhomogeneities in the disk.  For time scales of years, which
correspond to the dynamical
 time of the BLR,  one can invoke to account for these inhomogeneities, a structure like a
 spiral wave at $\sim 10$ light days from the center
 which would rotate at approximately the same
 speed as the disc, causing a
 shift of the peak in the profile in
 from 1000 to 2500~km/s in a few years (for a black hole mass of $7\times
10^7$Mo, the rotation timescale is on the order of 20 years). Such a
 single
 spiral arm is also invoked to explain the profiles and their variations
 in the LINER/Seyfert 1 galaxy NGC 1097 (Storchi-Bergman et al. 2003).
 A spiral wave will make a ``bump" in the
 disc, as it thicker than the rest of the disc. Therefore
 it would be more illuminated than the rest of the disc,
 causing a localized enhancement in the line profile.
  The rotation of the spiral wave should be accompanied by a change of the time lag
 between the continuum and the lines. In this case there is
no obvious link between a decrease of the continuum flux level and a velocity shift of the
 bump.

 There are however, other possible explanations different from that of a spiral arm.
 For instance one can invoke a  hot spot due to a collision of the disc with
a passing star (e.g. Syer, Clarke \& Rees ~\cite{sye}; Chakrabarti
\& Wiita~\cite{cha1993}), or the explosion of a supernova
releasing a remnant(a "SN cloud") (Collin \& Zahn 1999). In this case,
 the emitting
  material
 should be receding, with a velocity increasing with time.
In the model of the SN cloud, one can explain the correlated decrease
 of the continuum by assuming that as  the SN cloud  grows, it
 begins to cover the line of sight of the central source (in such a
 dense environment, a supernova would be denser, with a much larger column
 density than a normal supernova, during a few years).  In both cases the new material
 will be incorporated later on
 to the inner disc and the accretion rate will increase. But this would
 take place a long time after the explosion of the supernova or the
 capture of the star by the disc, in the
 viscous timescale of the inner disc, i.e. in tens of years, so we
 cannot check this hypothesis at present.

 For the two peak structure, one could invoke a double-armed spiral shock structure
as proposed by Chakrabarti and Wiita~(\cite{cha}).

 In a detailed physical study of the Balmer line emitting region in NGC 5548,
  Dumont et al. (1998)
  considered a sample of observations
 spanning a relatively small time (9 months) in 1989. They concluded that it
 must be
 heated partially by a non radiative mechanism, in particular because
 the variations of the lines were too small compared to those of the
 UV continuum in short term scales. Indeed, we noticed in
the present paper that the line flux variations are very small in short
 time scales. This non radiative heating can be due for instance to
 sound waves. Such waves can be easily
 provided by a spiral wave or by a supernova, or by a star colliding
with the disc.

 So, we can suggest that the BLR is a highly
 turbulent medium in rotation constituting a kind of semi-thick
 inhomogeneous disc. It is partly heated  by the central UV-X source,
 and partly mechanically through the dissipation of turbulent
 motions. At short time scales, instabilities of the inner disc create
 rapid changes of the continuum flux, inducing the reverberation
 effects, but smoothed by the non radiative part of the heating which
 varies much more slowly. At longer time scales, structural changes of
 the BLR induce changes of the line profiles and of the reverberated
 flux, and should lead after some years to changes in the accretion rate
 and consequently changes in the continuum flux.
 The inhomogeneities in the disc will boost the emissivity at
specific radial velocities, thus producing bumps or asymmetries
in the line profiles. However, one cannot choose among a variety
of models of inhomogeneous disc the one which suits better
the observed evolution of the Balmer line profiles in  NGC 5548 in 1996-2002.

The variations of the integral Balmer decrement can also be
interpreted in this framework. In Section ~\ref{sec4} (point 5),
we pointed out that the integrated Balmer decrement is
anticorrelated with the continuum flux level. As the line flux in
H$\alpha$ and H$\beta$ correlates well with that of the
continuum, we infer that the change of the integrated Balmer
decrement is also caused by changes in the continuum flux.
Indeed, when the ionization parameter decreases for a constant
density plasma, an increase of $F$(H$\alpha$)/$F$(H$\beta$)
intensity ratio is expected. (cf. for instance Wills et
al.~\cite{wil}). This is due to the decrease of the excitation
state of the ionized gas:  the temperature of the ionized zone
being smaller, the population of the upper levels with respect to
the lower ones decreases. However the great differences observed
in the behaviour of the Balmer decrement along the individual
profiles are certainly not due to the variations of the ionizing
flux, as they take place in longer time scales. They are more
probably due to local density fluctuations of the inhomogeneous
medium, leading also to variations of the ionization parameter,
and inducing changes of the Balmer decrement across the line
profiles.

\section{Conclusions}
\label{sec6}

Between  1996 and  2002, we have carried out a spectral
monitoring of the Seyfert galaxy NGC 5548. It is found that the
flux in the lines and the continuum gradually decreased and
reached minimum values in May-June 2002, i.e. the continuum
source gradually faded. The line wings at maximum light states
(1998) correspond to a Sy1 type, while at minimum light states
(2002), they are similar to a Sy1.8 type. It was shown that the
observed mean and rms line profiles, the variable part of the
H$\beta$ and H$\alpha$ profiles, and the difference profiles
present double peaks at radial velocities $\sim \pm 1000$~km/s.
During 1996, the red peak  was brighter than the blue one, and in
other years  (1998 -- 2002) the blue peak became brighter than
the red one. In 2000 -- 2002, we observed a third distinct peak
in the red wings of H$\alpha$ and H$\beta$, at a radial velocity
of $\sim +2500$~km/s. The radial velocity of this peak decreased
from 2000 to 2002 by $\sim 500$~km/s on the observed profiles,
i.e. this peak moved gradually across the line profile towards
lower radial velocities. The observed fluxes in  different parts
of the broad lines vary quasi-simultaneously. The flux in
different parts of the profiles, are well correlated with each
other, and with the continuum flux as well. Yet, the change in
shapes of the variuos parts of the lines, either they weakly
correlate, or simply do not correlate at all with the changes in
the continuum flux. Hence, these changes are not caused by
reverberation effects. We have observed a general increase of the
integral Balmer decrement with a continuum flux decrease.  This
probably due to a decrement in the ionization parameter, caused
by a decrease of the flux in the continuum.
 The behavior of the Balmer
decrement across the line profiles,  differs greatly
from time to time, is probably due to variations of density in
inhomogeneities in an accretion disc.

 Section ~\ref{sec5} tries to integrate all these observations in the
 framework of a model, and presents arguments in favor of the formation of
the
broad Balmer lines in a turbulent, partly mechanically heated
 accretion disc including large, moving
"thick" inhomogeneities, capable of  reprocessing the central source continuum.

\begin{acknowledgements}

The authors are grateful to Gaskell C.M. and Komberg B.V. for
useful discussions, Sergeev S.G. for providing a number of
service programs, Spangenberg L.I. for help in preparing the
paper, and Zhdanova V.E. for help in processing the spectra.

This paper has had financial support from INTAS (grant N96-0328),
RFBR (grants N97-02-17625, N00-02-16272 and N03-02-17123), state
program ``Astronomy'' (Russia), and CONACYT research grants
G28586-E, 32106-E, and 39560-F (Mexico).

\end{acknowledgements}

%\Online
\appendix\section{Modification of the spectrum scaling method}
\label{ap}

In the monitoring programs of AGN, the spectral scaling scheme
described by van Groningen \& Wanders (\cite{van}) is used. The
main idea of the algorithm is the creation of the difference
spectrum between an input  spectrum and a reference spectrum, for
which the flux is assumed to be  constant. The difference
spectrum is represented by the simple analytical function (usually
by a 2nd order polynomial). Then a $\chi^2$ of this correspondence
is minimized by a grid search method by successive variations of 3
input parameters: a flux scaling factor, a wavelength shift, and a
difference in resolution of the spectrum ($\Delta$ FWHM). For the
latter, a convolution with Gaussian function is selected. For data
with the S/N ratio $\sim 10$, the error of the flux scaling factor
is $<5$\%. However, the method of van Groningen \& Wanders
(\cite{van}) is unstable to the selection of initial parameters (
the zero approximation is done manually). In  order to circumvent
this problem, we have modified the method. The difference between
the individual spectra (obj) and the ``reference'' one (ref) is
represented by a 3rd degree polynomial, and for the minimization
of the differences, a downhill simplex method by Nelder and Mead
(\cite{nel}) is used.  The latter is more stable than the grid
search method used by van Groningen \& Wanders (\cite{van}). As a
zero approximation, the flux  in the spectrum lines was determined
automatically after subtraction of a linear continuum determined
by the beginning and the end of a  given spectral interval. The
scaling procedure is then carried out with the program means in
IDL, the program is fast and stable. The program output is similar
to that of van Groningen \& Wanders (\cite{van}): the flux scaling
factor, relative wavelength shift and  Gaussian width which is
used for convolution with one of the spectra for spectral
resolution correction, values of $\chi^2$ and  $\sigma$ for the
power approximation to the spectrum difference, the scaled and
difference spectra (obj-ref). The latter are obtained after
reducing the spectra to the same spectral  resolution. In order to
check the correctness of the scaling method several tests have
been carried out.

\begin{enumerate}
\item
We have tested the accuracy in the determination of the flux scaling
factor by means of a model spectrum of different S/N ratios.  As
model data, we adopted the result of a multi Gaussian approximation to
the Jan 21,1998  NGC~5548 spectrum, for the wavelength interval
(4700 -- 5400)\,\AA\AA\, including the
[\ion{O}{iii}]$\lambda\lambda$\,4959, 5007 and H$\beta$ emission
lines.

In testing, the oxygen line intensities remained constant while the
H$\beta$ intensity was varied between 10 and 500 percent (H$\beta= 100$\%
corresponded to the observed spectrum of NGC~5548 in Jan 21
1998). A continuum level was added to the lines and the spectrum
was convolved with grey noise, to a chosen value of the S/N ratio.
200 experiments were carried out for each of the selected S/N
ratios. From the simulated spectra with S/N=20, the average values of
the flux scaling factor show systematic decrease from 1.015 to
0.995 for changes of the H$\beta$ intensity from 10\% to 500\%.
The dispersion of the flux scaling factor being about 2.5\%
 As the S/N ratio is increased to 40 (typical value for
spectra obtained in our monitoring campaign), systematic errors
and dispersion values went down to 0.5\% and 1\% respectively.

\item
With the  model spectra we tested the method accuracy for the
determination of the flux scaling factor in the case in which
spectra of  different resolution are compared. In this case the
model spectrum adopted is the same as in case 1., with H$\beta$
intensity 100\%, but before introducing noise, a convolution with
a Gaussian of the proper width is done. The value of the Gaussian
dispersion ($\sigma$=FWHM/2.35) varied from 0.01 to 5.59 pixels
with steps of 0.01 pixel (dispersion being 2\,\AA/px). We found
that for spectra with S/N ratio $\sim 20$, the value of $\sigma
\geq 0.51$ and the flux scaling factor varies by less than 1\%.
We also checked  for changes of the flux scaling factor in the
observed blue spectra of NGC~5548. For this purpose, we first
scaled the spectra adopting a spectrum with a resolution ($\sim
8$\AA) as a referenced one. Then, the scaled spectra were
processed by the program adopting a spectrum of lower resolution
($\sim 15$\AA) as a the reference one. The values for the flux
scaling factors so derived, coincided within an accuracy of $\sim
1$\%.

\item
The wavelength shift ($\Delta \lambda$) is always recovered with
high accuracy by the method.

Thus  by scaling our AGN spectra by means of the modified method
of van Groningena \& Wanders (\cite{van}) as described above, one
can  obtain correct values of the scale parameters, their errors
being dependent only on the quality of the spectra.
\end{enumerate}
%\end{appendix}
\section{The tables presented in electronic form}
%\section{Appendix B:The tables presented in electronic form}
\label{ap_el}

\clearpage

\begin{onecolumn}

\setcounter{table}{1} \centering \tablecaption{Log of the spectroscopic
observations. Columns: 1 - number; 2 - UT date; 3 - Julian date; 4- code
according to Table 1; 5 - projected spectrograph entrance apertures; 6 -
wavelength range covered; 7 - spectral resolution; 8 - mean seeing; 9 -
position angle (PA) in degrees; 10 - signal to noise ratio in the continuum
(5160 -- 5220)\,\AA\AA\, near H$\beta$ and (6940 -- 7040)\,\AA\AA\, near
H$\alpha$.} \label{tab2a} \tablehead{ \hline \hline
  No &UT-Date   &    JD    &  Code& Aperture  &  Sp.range & Res. & Seeing & PA   & S/N \\
     &          &(2400000+)&      & (arcsec)  &    (\AA\AA)& (\AA)& arcsec& (deg)&5160-5220 \AA\AA \\
\hline
  1   &    2    & 3        &   4  & 5         &   6       & 7    & 8      & 9    & 10 \\
\hline}
\begin{supertabular}{lccccccccc}
  1 &1996Jan14 &50097.570 &  L1 & 2.4$\times$13.5 & 3600-7200 &  9   & 4  &     &  28 \\
  2 &1996Jan15 &50098.581 &  L1 & 4.2$\times$13.5 & 3600-7200 &  9   & 4  &     &  17 \\
  3 &1996Feb14 &50127.579 &  L  & 1.5$\times$6.0  & 3100-5800 &  6   & 2  &     &  95 \\
  4 &1996Feb14 &50127.569 &  L  & 1.5$\times$6.0  & 4500-7200 &  6   & 2  &     &  69 \\
  5 &1996Feb14 &50128.432 &  L  & 1.5$\times$6.0 & 3100-5700 &  6   &  3  &     & 114 \\
  6 &1996Feb14 &50128.472 &  L  & 1.5$\times$6.0  & 4500-7200 &  6  & 2  &     & 138 \\
  7 &1996Mar19 &50162.382 &  L  & 2.0$\times$6.0 & 3600-5600 &  8  & 3  &  40 &  83 \\
  8 &1996Mar19 &50162.399 &  L  & 2.0$\times$6.0  & 4700-7400 &  6  & 3  &  40 &  92 \\
  9 &1996Mar21 &50164.395 &  L  & 2.0$\times$6.0  & 3600-5600 &  8  & 3  &  45 &  79 \\
 10 &1996Mar21 &50164.412 &  L  & 2.0$\times$6.0 & 4700-7400 &  6  & 3  &  45 &  77 \\
 11 &1996Jul10 &50275.283 &  L  & 2.0$\times$6.0 & 3600-5400 &  8  & 1.4&  82 &  96 \\
 12 &1996Jul10 &50275.353 &  L  & 2.0$\times$6.0  & 5200-7000 &  8  & 1.4&  82 & 114 \\
 13 &1997Apr05 &50544.414 &  L1 & 4.2$\times$19.8 & 4100-5900 &  9  & 4  &  90 &  40 \\
 14 &1997Apr05 &50544.455 &  L1 & 4.2$\times$19.8 & 5500-7300 &  9  & 4  &  90 &  29 \\
 15 &1997Apr08 &50547.331 &  L  & 2.0$\times$6.0 & 4444-5244 &  4.5& 3  &   6 &  30 \\
 16 &1997Apr14 &50553.322 &  L  & 2.0$\times$6.0 & 4400-5300 &  4.5& 4  &   9 &  41 \\
 17 &1997Apr14 &50553.330 &  L  & 2.0$\times$6.0  & 6200-7100 &  4.5& 4  &   9 &  44 \\
 18 &1998Jan21 &50834.632 &  L  & 2.0$\times$6.0 & 3800-6200 &  8  & 3  &  62 &  97 \\
 19 &1998Jan21 &50834.639 &  L  & 2.0$\times$6.0  & 5900-7598 &  8  & 3  &  69 & 126 \\
 20 &1998Feb23 &50867.535 &  L  & 2.0$\times$6.0 & 3800-6200 &  8  & 2  &  53 &  73 \\
 21 &1998Feb23 &50867.539 &  L  & 2.0$\times$6.0  & 5900-7598 &  8  & 2  &  58 &  94 \\
 22 &1998Apr27 &50931.474 &  L1 & 8.0$\times$19.8 & 5500-7300 &  9  & 4  &     &  28 \\
 23 &1998Apr30 &50934.438 &  L1 & 8.0$\times$19.8 & 5500-7300 &  9  & 4  & 146 &  48 \\
 24 &1998Apr30 &50934.467 &  L1 & 8.0$\times$19.8 & 4000-5800 &  9  & 4  & 146 &  73 \\
 25 &1998May04 &50938.293 &  L  & 2.0$\times$6.0 & 3700-6200 &  8  & 2.5&  22 &  63 \\
 26 &1998May04 &50938.289 &  L  & 2.0$\times$6.0  & 5900-7598 &  8  & 2.5&  20 &  49 \\
 27 &1998May06 &50940.433 &  L  & 2.0$\times$6.0 & 3700-6200 &  8  & 3  & 108 &  62 \\
 28 &1998May06 &50940.436 &  L  & 2.0$\times$6.0  & 5900-7598 &  8  & 3  & 107 &  54 \\
 29 &1998May07 &50941.551 &  L  & 2.0$\times$6.0 & 3700-6200 &  8  & 3  &  93 & 106 \\
 30 &1998May08 &50942.452 &  L  & 2.0$\times$6.0 & 4100-5450 &  4.5& 2  & 104 &  41 \\
 31 &1998May08 &50942.461 &  L  & 2.0$\times$6.0 & 3700-6200 &  8  & 2  & 102 &  78 \\
 32 &1998May08 &50942.472 &  L  & 2.0$\times$6.0  & 5700-7700 &  8  & 2  & 100 &  22 \\
 33 &1998Jun17 &50982.407 &  L1 & 8.0$\times$19.8 & 4050-5850 &  9  & 2  &   0 &  74 \\
 34 &1998Jun19 &50984.287 &  L1 & 8.0$\times$19.8 & 4050-6147 &  9  & 4  &   0 &  81 \\
 35 &1998Jun19 &50984.328 &  L1 & 8.0$\times$19.8 & 5500-7300 &  9  & 4  &   0 &  47 \\
 36 &1998Jun20 &50985.350 &  L1 & 8.0$\times$19.8 & 4050-6000 &  9  & 2  &   0 &  67 \\
 37 &1998Jun20 &50985.281 &  L1 & 8.0$\times$19.8 & 5500-7300 &  9  & 2  &   0 &  32 \\
 38 &1998Jun25 &50990.409 &  L  & 2.0$\times$6.0 & 3600-6100 &  8  & 3  &  84 &  61 \\
 39 &1998Jun26 &50991.331 &  L  & 2.0$\times$6.0 & 3600-6100 &  8  & 2  & 102 &  40 \\
 40 &1998Jul14 &51008.778 &  GH & 2.5$\times$6.0 & 3979-7230 & 15  & 2  &  90 & 106 \\
 41 &1998Jul17 &51011.683 &  GH & 2.5$\times$6.0 & 4210-7471 & 15  & 2  &  90 &  84 \\
 42 &1998Jul26 &51020.703 &  GH & 2.5$\times$6.0 & 3960-7231 & 15  & 2.2&  90 &  73 \\
 43 &1998Jul27 &51021.712 &  GH & 2.5$\times$6.0 & 3930-7219 & 15  & 2.5&  90 &  85 \\
 44 &1999Jan13 &51191.952 &  GH & 2.5$\times$6.0 & 4150-7436 & 15  & 2.4&  90 &  94 \\
 45 &1999Jan14 &51193.011 &  GH & 2.5$\times$6.0 & 4151-7438 & 15  & 2  &  90 &  94 \\
 46 &1999Jan22 &51200.542 &  L1 & 4.2$\times$19.8 & 4000-5800 &  9  & 3  &  90 &  62 \\
 47 &1999Jan23 &51201.579 &  L1 & 4.2$\times$19.8 & 5550-7350 &  9  & 3  &  90 &  57 \\
 48 &1999Jan25 &51203.565 &  L1 & 4.2$\times$19.8 & 4000-5800 &  9  & 4  &  90 &  49 \\
 49 &1999Jan26 &51204.576 &  L1 & 4.2$\times$19.8 & 5550-7350 &  9  & 4  &  90 &  59 \\
 50 &1999Mar15 &51252.970 &  GH & 2.5$\times$6.0 & 4200-7500 & 15  & 2  &  90 & 138 \\
 51 &1999Mar24 &51262.444 &  L1 & 4.2$\times$19.8 & 4050-5850 &  9  & 3  &  90 &  68 \\
 52 &1999Mar24 &51262.452 &  L1 & 4.2$\times$19.8 & 5550-7350 &  9  & 3  &  90 &  15 \\
 53 &1999Mar24 &51262.452 &  L  & 2.0$\times$6.0 & 4260-5520 &  4.5& 3  &  89 &  93 \\
 54 &1999Apr11 &51279.533 &  L1 & 4.2$\times$19.8 & 4050-5850 &  9  & 4  &   0 &  59 \\
 55 &1999Jun12 &51342.409 &  L1 & 4.2$\times$19.8 & 3999-5796 &  9  & 3  &   0 &  29 \\
 56 &1999Jun14 &51344.354 &  L1 & 4.2$\times$19.8 & 4050-5850 &  9  & 3  &  90 &  63 \\
 57 &1999Jun16 &51346.335 &  L1 & 4.2$\times$19.8 & 4050-5850 &  9  & 3  &  90 &  21 \\
 58 &1999Jul16 &51376.277 &  L  & 2.0$\times$6.0 & 4000-5840 &  8  & 2  & 138 &  72 \\
 59 &1999Jul30 &51390.253 &  L1 & 4.2$\times$19.8 & 4000-5800 &  9  & 2  &   0 &  34 \\
 60 &1999Aug08 &51399.254 &  L1 & 4.2$\times$19.8 & 4000-5800 &  9  & 4  &   0 &  41 \\
 61 &1999Aug09 &51400.245 &  L1 & 4.2$\times$19.8 & 5550-7350 &  9  & 2  &   0 &  35 \\
 62 &2000Jan10 &51553.567 &  L1 & 4.2$\times$19.8 & 4000-5800 &  9  & 3  &  90 &  78 \\
 63 &2000Jan27 &51570.959 &  GH & 2.5$\times$6.0 & 4070-7350 & 15  & 2.5&  90 &  83 \\
 64 &2000Jan28 &51571.914 &  GH & 2.5$\times$6.0 & 4070-7350 & 15  & 1.5&  90 &  72 \\
 65 &2000Feb14 &51588.510 &  L1 & 4.2$\times$19.8 & 4020-5820 &  9  & 4  &  90 &  53 \\
 66 &2000Feb15 &51589.506 &  L1 & 4.2$\times$19.8 & 4030-5830 &  9  & 4  &  90 &  73 \\
 67 &2000Feb26 &51600.898 &  GH & 2.5$\times$6.0 & 4560-7850 & 15  & 2.2&  90 &  85 \\
 68 &2000Feb27 &51601.847 &  GH & 2.5$\times$6.0 & 4310-7600 & 15  & 2.5&  90 &  97 \\
 69 &2000Mar14 &51618.483 &  L1 & 4.2$\times$19.8 & 4050-5850 &  9  & 4  &  90 &  34 \\
 70 &2000Mar15 &51619.478 &  L1 & 4.2$\times$19.8 & 4050-5850 &  9  & 4  &  90 &  50 \\
 71 &2000Apr01 &51636.414 &  L1 & 4.2$\times$19.8 & 4100-5900 &  9  & 4  &  90 &  32 \\
 72 &2000Apr04 &51639.397 &  L1 & 4.2$\times$19.8 & 4050-5850 &  9  & 3  &  90 &  41 \\
 73 &2000Apr25 &51659.839 &  GH & 2.5$\times$6.0 & 4210-7460 & 15 &2.2 &  90 &  61 \\
 74 &2000Apr26 &51660.815 &  GH & 2.5$\times$6.0 & 4210-7460 & 15  & 2.2&  90 &  48 \\
 75 &2000May11 &51676.365 &  L1 & 4.2$\times$19.8 & 4050-5850&  9  & 3  &  90 &  48 \\
 76 &2000May16 &51681.283 &  L1 & 4.2$\times$19.8 & 5550-7350 &  9  & 3  &  90 &  52 \\
 77 &2000May20 &51685.321 &  L1 & 4.2$\times$19.8 & 5600-7400 &  9  & 2  &  90 &  61 \\
 78 &2000May21 &51686.258 &  L1 & 4.2$\times$19.8 & 4050-5850 &  9  & 3  &  90 &  53 \\
 79 &2000May21 &51686.350 &  L1 & 4.2$\times$19.8 & 5550-7350 &  9  & 2  &  90 &  41 \\
 80 &2000May25 &51689.786 &  GH & 2.5$\times$6.0  & 4130-7400 & 15  & 2.5&  90 &  96 \\
 81 &2000May26 &51690.835 &  GH & 2.5$\times$6.0 & 4134-7380 & 15  & 2.5&  90 &  71 \\
 82 &2000Jun06 &51702.378 &  L1 & 4.2$\times$19.8 & 4020-5820 &  9  & 3  &  90 &  56 \\
 83 &2000Jun08 &51704.299 &  L1 & 4.2$\times$19.8 & 4020-5820 &  9  & 3  &  90 &  47 \\
 84 &2000Jun15 &51711.308 &  L1 & 4.2$\times$19.8 & 5600-7400 &  9  & 3  &  90 &  41 \\
 85 &2000Jun25 &51721.712 &  GH & 2.5$\times$6.0 & 4690-7980 & 15  & 2.8&  90 &  49 \\
 86 &2000Jul12 &51738.335 &  L1 & 4.2$\times$19.8 & 5560-7360 &  9  & 3  &  90 &  33 \\
 87 &2000Jul13 &51739.275 &  L1 & 4.2$\times$19.8 & 5560-7360 &  9  & 3  &  90 &  44 \\
 88 &2000Jul28 &51754.269 &  L1 & 4.2$\times$19.8 & 4060-5860 &  9  & 3  &  90 &  38 \\
 89 &2000Jul29 &51755.301 &  L1 & 4.2$\times$19.8 & 4040-5840 &  9  & 2  &  90 &  46 \\
 90 &2000Jul3031$^{\mathrm{a}}$& 51756.786& L1 & 4.2$\times$19.8 & 4040-5840 &  9 & 2  &  90 &  90 \\
 91 &2001Jan25 &51935.482 &  L1 & 4.2$\times$19.8 & 4050-5850 &  9  & 2  &  90 &  77 \\
 92 &2001Jan26 &51936.517 &  L1 & 4.2$\times$19.8 & 5550-7350 &  9  & 2  &  90 &  58 \\
 93 &2001Jan29 &51938.560 &  L1 & 4.2$\times$19.8 & 4094-5740 &  9  & 2  &  90 &  58 \\
 94 &2001Feb03 &51944.498 &  L1 & 4.2$\times$19.8 & 4094-5740 &  9  & 2  &  90 &  54 \\
 95 &2001Feb16 &51957.492 &  L1 & 4.2$\times$19.8 & 4094-5740 &  9  & 2  &  90 &  52 \\
 96 &2001Mar13 &51981.867 &  GH & 2.5$\times$6.0 & 4120-7390 & 15  & 2.5&  90 &  78 \\
 97 &2001Apr13 &52013.360 &  L1 & 4.2$\times$19.8 & 4094-5740 &  9  & 3  &  90 &  69 \\
 98 &2001Apr13 &52013.451 &  L1 & 4.2$\times$19.8 & 5550-7350 &  9  & 3  &  90 &  50 \\
 99 &2001May05 &52034.784 &  GH & 2.5$\times$6.0 & 4156-7460 & 15  & 2  &  90 & 113 \\
100 &2001May12 &52041.829 &  GH & 2.5$\times$6.0 & 3600-6900 & 15  & 1.8&  90 &  96 \\
101 &2001May14 &52043.859 &  GH & 2.5$\times$6.0 & 3980-7300 & 15  & 1.8&  90 & 125 \\
102 &2001Jun14 &52074.772 &  GH & 2.5$\times$6.0 & 4022-7330 & 15  & 1.8&  90 &  93 \\
103 &2001Jun15 &52075.738 &  GH & 2.5$\times$6.0 & 4010-7330 & 15  & 1.8&  90 & 100 \\
104 &2001Jul10 &52101.367 &  L  & 2.0$\times$6.0 & 3630-6050 &  8  & 2  &  90 &  56 \\
105 &2001Jul10 &52101.367 &  L  & 2.0$\times$6.0 & 5740-8050 &  8  & 2  &  90 &  25 \\
106 &2001Jul12 &52103.283 &  L  & 2.0$\times$6.0 & 3630-6050 &  8  & 2  &   0 & 107 \\
107 &2001Jul12 &52103.283 &  L  & 2.0$\times$6.0 & 5740-8050 &  8  & 2  &   0 &  63 \\
108 &2001Jul21 &52112.340 &  L1 & 4.2$\times$19.8 & 4042-5800 &  9  & 2  &  90 &  34 \\
109 &2001Jul22 &52113.261 &  L  & 2.0$\times$6.0 & 3600-6020 &  8  & 2  &   0 &  92 \\
110 &2001Jul22 &52113.261 &  L  & 2.0$\times$6.0 & 5740-8050 &  8  & 2  &   0 &  46 \\
111 &2001Jul23 &52114.335 &  L1 & 8.0$\times$19.8 & 4094-5746 &  9  & 2  &  90 &  47 \\
112 &2001Jul26 &52117.333 &  L1 & 8.0$\times$19.8 & 4094-5746 &  9  & 2  &  90 &  62 \\
113 &2001Aug09 &52131.300 &  L1 & 4.2$\times$19.8 & 4400-6200 & 9  & 2  &  90 &  40 \\
114 &2002May15 &52410.456 &  L1 & 4.2$\times$19.8 & 4400-6200 &  9  & 2  &  90 &  52 \\
115 &2002May17 &52412.435 &  L1 & 4.2$\times$19.8 & 4400-6200 &  9  & 2  &  90 &  49 \\
116 &2002Jun04 &52430.735 &  GH & 2.5$\times$6.0  & 3976-7305 & 15   & 2  &  90 & 114 \\
\hline
\end{supertabular}
\begin{list}{}{}
\item[$^{\mathrm{a}}$] Date 2000 Jul3031 means average of data during two nights;
for this case average JD is given.
\end{list}

\clearpage

\setcounter{table}{4} \centering \tablecaption{Observed continuum, H$_\beta$
and H$_\alpha$ fluxes. Columns: 1 - UT-Date; 2 - Julian date; 3\,- a telescope
code, according to Table~\ref{tab1}; 4 - $F$(cont), the continuum flux at
5190\,\AA\, (in units of $10^{-15}$\, erg\,s$^{-1}$\,cm$^{-2}$\,\AA$^{-1}$),
reduced to the 1 m telescope aperture $4.2''\times 19.8''$; 5 -
$\varepsilon_c$, the estimated continuum flux error; 6 - $F$(H$\beta$), the
H$\beta$ total flux (in units of 10$^{-13}$\,erg\,s$^{-1}$\,cm$^{-2}$); 7 -
$\varepsilon_{{\rm H}\beta}$, the H$\beta$ flux error; 8\,- $F$(H$\alpha$), the
H$\alpha$ total flux (in units of $10^{-13}$\,erg\,s$^{-1}$\,cm$^{-2}$); 9 -
$\varepsilon_{{\rm H}\alpha}$, the H$\alpha$ flux error.} \label{tab5a}
\tablehead{ \hline \hline
UT-Date & JD & Code& F(cont) & $\varepsilon$ & F(H$_\beta$)& $\varepsilon$    & F(H$_\alpha$) &$ \varepsilon$ \\
        &(2400000+)&         & (5190)\AA     &             &(4795-5018)\AA\AA &               &(6500-6800)\AA\AA &             \\
\hline
  1     &    2      & 3 &   4   & 5  &   6  & 7 & 8 & 9  \\
\hline}
\begin{supertabular}{lcccccccc}
 1996Jan14 &50097.570 & L1 &       &      &  7.32 & 0.22 &   29.90 & 1.92   \\
 1996Jan15 &50098.581 & L1 &       &      &       &      &   32.74 & 2.10   \\
 1996Feb14 &50127.579 & L  &       &      &  8.26 & 0.24 &         &        \\
 1996Feb14 &50127.569 & L  &       &      &       &      &   29.76 & 1.98   \\
 1996Feb14 &50128.432 & L  &       &      &  8.79 & 0.25 &         &        \\
 1996Feb14 &50128.472 & L  &       &      &       &      &   32.81 & 2.18   \\
 1996Mar19 &50162.382 & L  &  14.13& 1.03 &  7.85 & 0.08 &         &        \\
 1996Mar19 &50162.399 & L  &       &      &       &      &   28.47 & 2.59   \\
 1996Mar21 &50164.395 & L  &  12.74& 0.93 &  7.89 & 0.08 &         &        \\
 1996Mar21 &50164.412 & L  &       &      &       &      &   25.04 & 2.28   \\
 1996Jul10 &50275.283 & L  &  16.98& 0.51 & 10.04 & 0.30 &         &        \\
 1996Jul10 &50275.353 & L  &       &      &       &      &   41.38 & 2.07   \\
 1997Apr05 &50544.414 & L1 &  10.83& 0.63 &  7.67 & 0.38 &         &        \\
 1997Apr05 &50544.455 & L1 &       &      &       &      &   36.69 & 1.83   \\
 1997Apr08 &50547.331 & L  &  11.75& 0.68 &  7.14 & 0.36 &         &        \\
 1997Apr14 &50553.322 & L  &  10.35& 0.31 &  7.53 & 0.23 &         &        \\
 1997Apr14 &50553.330 & L  &       &      &       &      &   37.53 &  .88   \\
 1998Jan21 &50834.632 & L  &  14.96& 0.45 &  8.96 & 0.27 &         &        \\
 1998Jan21 &50834.639 & L  &       &      &       &      &   34.65 & 1.73   \\
 1998Feb23 &50867.535 & L  &  14.36& 0.43 & 10.16 & 0.30 &         &        \\
 1998Feb23 &50867.539 & L  &       &      &       &      &   37.81 & 1.89   \\
 1998Apr27 &50931.474 & L1 &       &      &       &      &   42.10 & 1.68   \\
 1998Apr30 &50934.438 & L1 &       &      &       &      &  39.81:$^{\mathrm{b}} $ &
2.00   \\
 1998Apr30 &50934.467 & L1 &  18.49& 0.55 &  9.98 & 0.30 &         &        \\
 1998May04 &50938.289 & L  &       &      &       &      &   36.98 & 1.85   \\
 1998May04 &50938.293 & L  &  18.85& 0.83 & 10.57 & 0.12 &         &        \\
 1998May06 &50940.433 & L  &  20.05& 0.88 & 10.4  & 0.52 &         &        \\
 1998May06 &50940.436 & L  &       &      &       &      &   46.64:& 5.6    \\
 1998May07 &50941.551 & L  &  18.92& 0.78 & 11.16 & 0.28 &         &        \\
 1998May08 &50942.452 & L  &  18.06& 0.74 & 11.72 & 0.29 &         &        \\
 1998May08 &50942.461 & L  &  17.42& 0.71 & 11.34 & 0.28 &         &        \\
 1998May08 &50942.472 & L  &       &      &       &      &   39.60 & 1.98   \\
 1998Jun17 &50982.407 & L1 &  17.52& 0.18 & 11.42 & 0.23 &         &        \\
 1998Jun19 &50984.287 & L1 &  17.26& 0.17 & 11.74 & 0.23 &         &        \\
 1998Jun19 &50984.328 & L1 &       &      &       &      &   44.45 & 1.68   \\
 1998Jun20 &50985.281 & L1 &       &      &       &      &   47.31 & 1.79   \\
 1998Jun20 &50985.350 & L1 &  18.44&:0.85 & 11.55 & 0.14 &         &        \\
 1998Jun25 &50990.409 & L  &  16.41& 0.10 & 11.90 & 0.17 &         &        \\
 1998Jun26 &50991.331 & L  &  16.28& 0.10 & 11.67 & 0.16 &         &        \\
 1998Jul14 &51008.778 & GH &  15.25& 0.14 & 10.24 & 0.19 &   42.18 & 0.81   \\
 1998Jul17 &51011.683 & GH &  15.05& 0.14 & 10.52 & 0.20 &   43.35 & 0.84   \\
 1998Jul26 &51020.703 & GH &  15.25& 0.34 & 10.68 & 0.31 &   41.69 & 3.62   \\
 1998Jul27 &51021.712 & GH &  15.73& 0.35 & 10.26 & 0.30 &   36.86 & 3.20   \\
 1999Jan13 &51191.952 & GH &  13.78& 0.21 &  9.41 & 0.10 &   32.40 & 1.04   \\
 1999Jan14 &51193.011 & GH &  14.08& 0.21 &  9.34 & 0.10 &   33.90 & 1.08   \\
 1999Jan22 &51200.542 & L1 &  13.46& 0.40 &  9.44 & 0.28 &         &        \\
 1999Jan23 &51201.579 & L1 &       &      &       &      &   34.24 & 1.10   \\
 1999Jan25 &51203.565 & L1 &  14.26& 0.43 &  9.62 & 0.29 &         &        \\
 1999Jan26 &51204.576 & L1 &       &      &       &      &   34.45 & 1.10   \\
 1999Mar15 &51252.970 & GH &  15.63& 0.47 &  8.57 & 0.26 &   38.69:& 1.93   \\
 1999Mar24 &51262.444 & L1 &  16.59& 0.16 & 10.03 & 0.16 &         &        \\
 1999Mar24 &51262.452 & L1 &       &      &       &      &   36.17:& 1.81   \\
 1999Mar24 &51262.452 & L  &  16.53& 0.16 &  9.80 & 0.16 &         &        \\
 1999Apr11 &51279.533 & L1 &  15.10& 0.45 &  9.66 & 0.29 &         &        \\
 1999Jun12 &51342.409 & L1 &  12.55& 0.25 &  9.76 & 0.10 &         &        \\
 1999Jun14 &51344.354 & L1 &  12.66& 0.25 &  9.89 & 0.10 &         &        \\
 1999Jun16 &51346.335 & L1 &  13.00& 0.26 &  9.81 & 0.10 &         &        \\
 1999Jul16 &51376.277 & L  &  14.06& 0.42 & 10.12 & 0.30 &         &        \\
 1999Jul30 &51390.253 & L1 &  10.98& 0.33 &  8.36 & 0.25 &         &        \\
 1999Aug08 &51399.254 & L1 &  10.43& 0.31 &  8.76 & 0.26 &         &        \\
 1999Aug09 &51400.245 & L1 &       &      &       &      &   35.52 & 1.78   \\
 2000Jan10 &51553.567 & L1 &  10.03& 0.30 &  9.32 & 0.28 &         &        \\
 2000Jan27 &51570.959 & GH &  10.47& 0.31 &  7.95 & 0.26 &   33.45 & 2.16   \\
 2000Jan28 &51571.914 & GH &  12.52& 0.38 &  7.59 & 0.25 &   36.64 & 2.36   \\
 2000Feb14 &51588.510 & L1 &  10.55& 0.50 &  7.21 & 0.17 &         &        \\
 2000Feb15 &51589.506 & L1 &  11.27& 0.53 &  6.97 & 0.17 &         &        \\
 2000Feb26 &51600.898 & GH &   9.22& 0.13 &  6.34 & 0.20 &   29.62 & 2.24   \\
 2000Feb27 &51601.847 & GH &   9.40& 0.13 &  6.63 & 0.21 &   26.60 & 2.02   \\
 2000Mar14 &51618.483 & L1 &  11.10& 0.33 &  7.03 & 0.56 &         &        \\
 2000Mar15 &51619.478 & L1 &   9.98& 0.30 &  6.26 & 0.50 &         &        \\
 2000Apr01 &51636.414 & L1 &   8.82& 0.12 &  6.60 & 0.17 &         &        \\
 2000Apr04 &51639.397 & L1 &   8.65& 0.12 &  6.36 & 0.17 &         &        \\
 2000Apr25 &51659.839 & GH &   7.83& 0.08 &  5.68 & 0.06 &   28.15 & 1.64   \\
 2000Apr26 &51660.815 & GH &   7.73& 0.08 &  5.63 & 0.06 &   25.95 & 1.52   \\
 2000May11 &51676.365 & L1 &   8.59& 0.26 &  5.22 & 0.16 &         &        \\
 2000May16 &51681.283 & L1 &       &      &       &      &   22.84 & 1.14   \\
 2000May20 &51685.321 & L1 &       &      &       &      &   21.94 & 0.21   \\
 2000May21 &51686.258 & L1 &   8.21& 0.25 &  5.13 & 0.15 &         &        \\
 2000May21 &51686.350 & L1 &       &      &       &      &   22.24 & 0.21   \\
 2000May25 &51689.786 & GH &   8.13& 0.08 &  5.35 & 0.18 &   22.54 & 0.72   \\
 2000May26 &51690.835 & GH &   8.12& 0.08 &  5.07 & 0.15 &   23.59 & 1.18   \\
 2000Jun06 &51702.378 & L1 &   7.47& 0.22 &  5.65 & 0.18 &         &        \\
 2000Jun08 &51704.299 & L1 &   9.31& 0.28 &  5.40 & 0.17 &         &        \\
 2000Jun15 &51711.308 & L1 &       &      &       &      &   19.10 & 0.96   \\
 2000Jun25 &51721.712 & GH &   8.09& 0.24 &  5.07 & 0.15 &   18.09 & 0.90   \\
 2000Jul12 &51738.335 & L1 &       &      &       &      &   20.26 & 0.35   \\
 2000Jul13 &51739.275 & L1 &       &      &       &      &   19.76 & 0.35   \\
 2000Jul28 &51754.269 & L1 &   9.14& 0.27 &  6.37 & 0.07 &         &        \\
 2000Jul29 &51755.301 & L1 &   9.54& 0.29 &  6.47 & 0.07 &         &        \\
 2000Jul3031$^{\mathrm{a}}$& 51756.786 &L1&  9.01 & 0.36 &  6.23  & 0.17& & \\
 2001Jan25 &51935.482 & L1 &   8.53& 0.26 &  5.85 & 0.18 &         &        \\
 2001Jan26 &51936.517 & L1 &       &      &       &      &   27.50 & 1.38   \\
 2001Jan29 &51938.560 & L1 &   9.58& 0.29 &  5.86 & 0.18 &         &        \\
 2001Feb03 &51944.498 & L1 &   8.98& 0.27 &  5.90 & 0.18 &         &        \\
 2001Feb16 &51957.492 & L1 &   8.68& 0.26 &  6.28 & 0.19 &         &        \\
 2001Mar13 &51981.867 & GH &   9.09& 0.27 &  5.52 & 0.17 &   23.26 &  1.16  \\
 2001Apr13 &52013.360 & L1 &   8.85& 0.27 &  4.28 & 0.13 &         &        \\
 2001Apr13 &52013.451 & L1 &       &      &       &      &   21.00 &  1.05  \\
 2001May05 &52034.784 & GH &   7.71& 0.23 &  3.52 & 0.11 &   17.85 &  0.89  \\
 2001May12 &52041.829 & GH &   7.98& 0.08 &  3.32 & 0.04 &   14.97 &  0.72  \\
 2001May14 &52043.859 & GH &   7.94& 0.08 &  3.34 & 0.04 &   16.01 &  0.77  \\
 2001Jun14 &52074.772 & GH &   9.91& 0.20 &  4.55 & 0.08 &   18.57 &  0.82  \\
 2001Jun15 &52075.738 & GH &   9.64& 0.19 &  4.44 & 0.07 &   17.45 &  0.77  \\
 2001Jul10 &52101.367 & L  &  11.27& 0.11 &  6.36 & 0.22 &         &        \\
 2001Jul10 &52101.383 & L  &       &      &       &      &   26.01 &  0.38  \\
 2001Jul12 &52103.273 & L  &       &      &       &      &   26.56 &  0.39  \\
 2001Jul12 &52103.283 & L  &  11.31& 0.11 &  6.68 & 0.23 &         &        \\
 2001Jul21 &52112.340 & L1 &  11.45& 0.41 &  6.27 & 0.16 &         &        \\
 2001Jul22 &52113.261 & L  &  10.65& 0.38 &  6.56 & 0.17 &         &        \\
 2001Jul22 &52113.286 & L  &       &      &       &      &   24.13 &  1.21  \\
 2001Jul23 &52114.335 & L1 &  11.13& 0.40 &  6.28 & 0.16 &         &        \\
 2001Jul26 &52117.333 & L1 &  10.72& 0.30 &  6.49 & 0.15 &         &        \\
 2001Aug09 &52131.300 & L1 &  10.54& 0.32 &  4.95 & 0.15 &         &        \\
 2002May15 &52410.456 & L1 &   7.32& 0.22 &  2.64 & 0.08 &         &        \\
 2002May17 &52412.435 & L1 &   8.11& 0.24 &  2.64 & 0.08 &         &        \\
 2002Jun04 &52430.735 & GH &   7.41& 0.22 &  2.42 & 0.07 &   13.70 & 0.68   \\
\hline \multicolumn{9}{l}{\rule{0pt}{11pt}}
\end{supertabular}
%\end{tabular}
\begin{list}{}{}
\item[$^{\mathrm{a}}$] Date 2000 Jul3031 means average of data during two nights; for
this case average JD is given.
\item[$^{\mathrm{b}}$] Colon marks unsure value.
\end{list}
\end{onecolumn}

%\end{appendix}

%\listofobjects
\end{document}